\documentclass[12pt,a4paper]{article}
\usepackage{psfrag}
\usepackage{graphics}
\usepackage{amssymb,epsfig,amsmath,euscript,array}
\usepackage{cite}

\makeatletter
\@addtoreset{equation}{section}
\makeatother
\renewcommand{\theequation}{\thesection.\arabic{equation}}

\newcommand{\startappendix}{
\setcounter{section}{0}
\renewcommand{\thesection}{\Alph{section}}
\renewcommand{\theequation}{\Alph{section}.\arabic{equation}}}

\newcommand{\Appendix}[1]{
\refstepcounter{section}
\begin{flushleft}
{\Large\bf Appendix \thesection: #1}
\end{flushleft}}

\newcounter{multieqs}

\newcommand{\be}{\begin{equation}}
\newcommand{\ee}{\end{equation}}

\newcommand{\bra}[1]{\langle #1|}
\newcommand{\ket}[1]{|#1 \rangle}

\newcommand{\bm}[1]{\mbox{\boldmath $#1$}}

\def\bd{\begin{document}}
\def\ed{\end{document}}
\def\nn{\nonumber}
\def\bea{\begin{eqnarray}}
\def\eea{\end{eqnarray}}
\let\bm=\bibitem
\let\la=\label

\def\npb#1#2#3{Nucl. Phys. {\bf{B#1}} #3 (#2)}
\def\plb#1#2#3{Phys. Lett. {\bf{#1B}} #3 (#2)}
\def\prl#1#2#3{Phys. Rev. Lett. {\bf{#1}} #3 (#2)}
\def\prd#1#2#3{Phys. Rev. {D \bf{#1}} #3 (#2)}
\def\cmp#1#2#3{Comm. Math. Phys. {\bf{#1}} #3 (#2)}
\def\cqg#1#2#3{Class. Quantum Grav. {\bf{#1}} #3 (#2)}
\def\nppsa#1#2#3{Nucl. Phys. B (Proc. Suppl.) {\bf{#1A}}#3 (#2)}
\def\ap#1#2#3{Ann. of Phys. {\bf{#1}} #3 (#2)}
\def\ijmp#1#2#3{Int. J. Mod. Phys. {\bf{A#1}} #3 (#2)}
\def\rmp#1#2#3{Rev. Mod. Phys. {\bf{#1}} #3 (#2)}
\def\mpla#1#2#3{Mod. Phys. Lett. {\bf A#1} #3 (#2)}
\def\jhep#1#2#3{J. High Energy Phys. {\bf #1} #3 (#2)}
\def\atmp#1#2#3{Adv. Theor. Math. Phys. {\bf #1} #3 (#2)}

\newcommand{\EQ}[1]{\begin{equation} #1 \end{equation}}
\newcommand{\AL}[1]{\begin{subequations}\begin{align} #1 \end{align}\end{subequations}}
\newcommand{\SP}[1]{\begin{equation}\begin{split} #1 \end{split}\end{equation}}
\newcommand{\ALAT}[2]{\begin{subequations}\begin{alignat}{#1} #2 \end{alignat}\end{subequations}}
\def\beqa{\begin{eqnarray}}
\def\eeqa{\end{eqnarray}}
\def\beq{\begin{equation}}
\def\eeq{\end{equation}}

\def\N{{\cal N}}
\def\sst{\scriptscriptstyle}
\def\thetabar{\bar\theta}
\def\Tr{{\rm Tr}}
\def\one{\mbox{1 \kern-.59em {\rm l}}}
 \def\Nh{\hat{N}}

\def\a{\alpha}      \def\da{{\dot\alpha}}
\def\b{\beta}       \def\db{{\dot\beta}}
\def\c{\gamma}  \def\G{\Gamma}  \def\cdt{\dot\gamma}
\def\d{\delta}  \def\D{\Delta}  \def\ddt{\dot\delta}
\def\e{\epsilon}        \def\vare{\varepsilon}
\def\f{\phi}    \def\F{\Phi}    \def\vvf{\f}
\def\h{\eta}
\def\k{\kappa}
\def\l{\lambda} \def\L{\Lambda}
\def\m{\mu} \def\n{\nu}
\def\o{\omega}
\def\p{\pi} \def\P{\Pi}
\def\r{\rho}
\def\s{\sigma}  \def\S{\Sigma}
\def\t{\tau}
\def\th{\theta} \def\Th{\Theta} \def\vth{\vartheta}
\def\X{\Xeta}
\def\z{\zeta}

\def\cA{{\cal A}} \def\cB{{\cal B}} \def\cC{{\cal C}}
\def\cD{{\cal D}} \def\cE{{\cal E}} \def\cF{{\cal F}}
\def\cG{{\cal G}} \def\cH{{\cal H}} \def\cI{{\cal I}}
\def\cJ{{\cal J}} \def\cK{{\cal K}} \def\cL{{\cal L}}
\def\cM{{\cal M}} \def\cN{{\cal N}} \def\cO{{\cal O}}
\def\cP{{\cal P}} \def\cQ{{\cal Q}} \def\cR{{\cal R}}
\def\cS{{\cal S}} \def\cT{{\cal T}} \def\cU{{\cal U}}
\def\cV{{\cal V}} \def\cW{{\cal W}} \def\cX{{\cal X}}
\def\cY{{\cal Y}} \def\cZ{{\cal Z}}

\def\ua{\underline{\alpha}}
\def\ub{\underline{\phantom{\alpha}}\!\!\!\beta}
\def\uc{\underline{\phantom{\alpha}}\!\!\!\gamma}
\def\um{\underline{\mu}}
\def\ud{\underline\delta}
\def\ue{\underline\epsilon}
\def\una{\underline a}\def\unA{\underline A}
\def\unb{\underline b}\def\unB{\underline B}
\def\unc{\underline c}\def\unC{\underline C}
\def\und{\underline d}\def\unD{\underline D}
\def\une{\underline e}\def\unE{\underline E}
\def\unf{\underline{\phantom{e}}\!\!\!\! f}\def\unF{\underline F}
\def\unm{\underline m}\def\unM{\underline M}
\def\unn{\underline n}\def\unN{\underline N}
\def\unp{\underline{\phantom{a}}\!\!\! p}\def\unP{\underline P}
\def\unq{\underline{\phantom{a}}\!\!\! q}
\def\unQ{\underline{\phantom{A}}\!\!\!\! Q}
\def\unH{\underline{H}}

\def\As {{A \hspace{-6.4pt} \slash}\;}
\def\bs {{b \hspace{-6.4pt} \slash}\;}
\def\Ds {{D \hspace{-6.4pt} \slash}\;}
\def\ds {{\del \hspace{-6.4pt} \slash}\;}
\def\ss {{\s \hspace{-6.4pt} \slash}\;}
\def\ks {{ k \hspace{-6.4pt} \slash}\;}
\def\ps {{p \hspace{-6.4pt} \slash}\;}
\def\pas {{{p_1} \hspace{-6.4pt} \slash}\;}
\def\pbs {{{p_2} \hspace{-6.4pt} \slash}\;}

\def\Fh{\hat{F}}
\def\Vh{\hat{V}}
\def\Xh{\hat{X}}
\def\ah{\hat{a}}
\def\xh{\hat{x}}
\def\yh{\hat{y}}
\def\ph{\hat{p}}
\def\xih{\hat{\xi}}

\def\psit{\tilde{\psi}}
\def\Psit{\tilde{\Psi}}
\def\tht{\tilde{\th}}

\def\At{\tilde{A}}
\def\Qt{\tilde{Q}}
\def\Rt{\tilde{R}}
\def\Nt{\tilde{N}}

\def\at{\tilde{a}}
\def\st{\tilde{s}}
\def\ft{\tilde{f}}
\def\pt{\tilde{p}}
\def\qt{\tilde{q}}
\def\vt{\tilde{v}}
\def\nt{\tilde{n}}

\def\delb{\bar{\partial}}
\def\bz{\bar{z}}
\def\bD{\bar{D}}
\def\bB{\bar{B}}

\def\bk{{\bf k}}
\def\bl{{\bf l}}
\def\bp{{\bf p}}
\def\bq{{\bf q}}
\def\br{{\bf r}}
\def\bx{{\bf x}}
\def\by{{\bf y}}
\def\bR{{\bf R}}
\def\bV{{\bf V}}

\def\d{\delta}\def\D{\Delta}\def\ddt{\dot\delta}

\def\pa{\partial} \def\del{\partial}
\def\xx{\times}
\def\uno{\mbox{1 \kern-.59em {\rm l}}}

\def\trp{^{\top}}
\def\inv{^{-1}}
\def\dag{{^{\dagger}}}
\def\pr{^{\prime}}

\def\rar{\rightarrow}
\def\lar{\leftarrow}
\def\lrar{\leftrightarrow}

\newcommand{\0}{\,\!}      
\def\one{1\!\!1\,\,}
\def\im{\imath}
\def\jm{\jmath}

\newcommand{\tr}{\mbox{tr}}
\newcommand{\slsh}[1]{/ \!\!\!\! #1}

\def\vac{|0\rangle}
\def\lvac{\langle 0|}

\def\hlf{\frac{1}{2}}
\def\ove#1{\frac{1}{#1}}

\def\Box{\square}
\def\ZZ{\mathbb{Z}}
\def\CC#1{({\bf #1})}
\def\bcomment#1{}
\def\bfhat#1{{\bf \hat{#1}}}
\def\VEV#1{\left\langle #1\right\rangle}

\newcommand{\ex}[1]{{\rm e}^{#1}} \def\ii{{\rm i}}

\def\rr{{\rm r}} \def\rs{{\rm s}}\def\rv{{\rm v}}
\def\ri{{\rm i}}\def\rj{{\rm j}}

\newcommand{\lrbrk}[1]{\left(#1\right)}
\newcommand{\sfrac}[2]{{\textstyle\frac{#1}{#2}}}

\font\mybb=msbm10 at 12pt
\def\bb#1{\hbox{\mybb#1}}

\font\myBB=msbm10 at 18pt
\def\BB#1{\hbox{\myBB#1}}

\begin{document}

\hfill{ hep-th/0403188}

\vspace{20pt}

\begin{center}

{\Large \bf Fermion BMN operators, the dilatation operator \\}
\vspace{8pt}
{\Large \bf of $\cN=4$  SYM,
   and pp-wave string interactions\\}

\vspace{30pt}

{\bf George Georgiou$^{a}$ and Gabriele Travaglini$^{b}$ }

{\small \em
\begin{itemize}
\item[\ \ \ \ \ \ $^a$]
Centre for Particle Theory,
Department of Physics and IPPP, University of Durham, \\
Durham, DH1 3LE, UK
\item[\ \ \ \ \ \ $^b$]
Department of Physics, Queen Mary College,
London E1 4NS, UK
\end{itemize}
}


\vspace{10pt}

{\sffamily \tt george.georgiou@durham.ac.uk,
g.travaglini@qmul.ac.uk }

\vspace{30pt}
{\bf Abstract}

\end{center}

The goal of this paper is to study  the BMN correspondence
in the fermionic sector.
On the field theory side, we compute matrix elements
of the dilatation operator in $\cN =4$ Super Yang-Mills 
for BMN operators containing two fermion impurities.
Our calculations are  performed up to
and including $\cO (\lambda')$
in the  't Hooft coupling and $ \cO ( g_{2} )$
in the  Yang-Mills genus counting parameter.
On the string theory side, we compute the corresponding
matrix elements of the interacting string Hamiltonian
in string field theory, using the three-string
interaction vertex constructed  by Spradlin and Volovich
(and subsequently  elaborated by Pankiewicz and Stefanski).
In string theory we use the natural string basis,
and in field theory the basis which is isomorphic to it.
We find that the matrix elements computed in field theory and the
corresponding string amplitudes derived from
the three-string vertex are, in all cases,
in perfect agreement.

\vspace{0.5cm}

\setcounter{page}{0}
\thispagestyle{empty}
\newpage

\section{Introduction}
In \cite{BMN}, Berenstein, Maldacena and Nastase
(BMN) proposed an intriguing correspondence 
between type IIB superstring theory on a
pp-wave background geometry and a sector of $\N = 4$
Super Yang-Mills (SYM). 
BMN compared the exact expression \cite{Metsaev:2001bj} 
for the mass spectrum of the string states in free string theory,
$g_{\rm st}=0$, 
to the planar anomalous dimension of certain field theory operators
- since then called the BMN operators -
to the first order in the 't Hooft coupling of the theory 
$\l'$,
finding remarkable agreement.
Shortly after it was shown in \cite{zanon} that, 
thanks to the superconformal invariance of the $\cN=4$ theory, 
it was actually possible to reproduce from field theory
the full (all orders in $\lambda'$) 
expression for the masses of string states   at $g_{\rm st}=0$.
An important step forward was subsequently
taken in \cite{ver,gross2},
where the correspondence was expressed as
\cite{gross2}
\be
\label{ham}
{1\over \mu} \, {H}_{\rm string} = \Delta - J \ ,
 \ee
where ${H}_{\rm string}$ is the 
{\it interacting} string Hamiltonian, 
$\mu$ is the scale parameter
of the pp-wave metric, and $\Delta-J$ is
the difference between the gauge theory
dilatation operator $\Delta$ 
and  the R-charge $J$.
The relation \eqref{ham} is conjectured to
hold  in the double-scaling limit $N \sim J^2 \to \infty$
to all orders in the two parameters of the theory,
$g_2$ and $\l'$, where
\beqa
\label{lampr}
\l' &=& \frac{g_{\rm YM}^2 N}{J^2} \ = \
\frac{1}{(\mu p^+ \a')^2}\ ,
\\ \cr
\label{gtwo}
g_2 &=& \frac{J^2}{N}\ = \ 4 \pi \,g_{\rm st}\, (\mu p^+ \a')^2 \ .
\eeqa
Here $\l'$ is the effective 't Hooft coupling
of the BMN sector \cite{BMN}, and $g_2$ is the genus
counting parameter of Feynman diagrams
\cite{BMN,KPSS,Constable1}.
The right hand sides of \eqref{lampr}, \eqref{gtwo}
express $\l'$ and $g_2$ in terms
of the parameters in  pp-wave string theory, so that
$1/\l' \propto \mu$ measures the deviation from  flat
space $\mu \to 0$ and, importantly,
the Yang-Mills genus counting parameter  $g_2$
is proportional to the string coupling $g_{\rm st}$.

Tests of the relation  \eqref{ham} rely on the careful comparison
of string amplitudes obtained in pp-wave string field theory
to the matrix elements of the dilatation operator 
$\Delta$ in Yang-Mills, and have been performed, 
for the bosonic sector of the theory,
in a variety  of cases:
\vspace{-0.5cm}
\begin{itemize}
\item[{\bf a.}] in \cite{Gomis}, the case of BMN operators
with two scalar impurities of different flavour
was studied;
\item[{\bf b.}] in \cite{Gomis2,GKT}, 
BMN operators with an  arbitrary number of scalar impurities
was considered; and finally,
\item[{\bf c.}]
in \cite{GKT}, all the $SO(4) \times SO(4)$
representations of
two scalar impurity and two vector impurity
BMN operators were studied, as well as
BMN operators with mixed (one scalar/one vector) impurities.
\end{itemize}
In all cases, precise agreement was found between 
the string amplitudes 
obtained using the superstring vertex%
\footnote{
The  expression for this vertex was originally 
obtained by Spradlin and
Volovich \cite{SV1,SV2},
and further  studied and clarified by 
Pankiewicz and Stefanski  in 
\cite{AP,PS}.}
and 
the corresponding matrix elements obtained in  field theory
\cite{Gomis,Gomis2,GKT}. 
In particular, the analysis of \cite{GKT} clarified a puzzle
concerning the realisation 
of the $\bb{Z}_2\subset SO(8)$ symmetry of the 
pp-wave background geometry.%
\footnote{The presence of a
five-form flux in the pp-wave background breaks the 
lightcone $SO(8)$ symmetry down to $SO(4)\times SO(4)
\times \bb{Z}_2$. The
first (second)  $SO(4)$  rotates the first (last) four directions
$\{x^i \}$  
($\{x^a\}$),  while the $\bb{Z}_2$ 
symmetry swaps $\{x^i\} \to \{x^a \}$.} 
Apparently,
the $\bb{Z}_2$ part of the bosonic symmetry of the pp-wave
background is not respected by the Spradlin-Volovich
three-string interactions 
\cite{KKLP,LMP,SV2,CKPRT,KKPR}:
a relative minus sign appears in the string amplitude 
involving states with two oscillators along the first $SO(4)$
compared to that with two oscillators along the second $SO(4)$.
However, the string vertex is invariant under $\bb{Z}_2$, 
and the puzzle is solved,  if one makes the 
parity assignment
under the $\bb{Z}_2$ symmetry:
$\ket{0} \rightarrow \ket{0}$,
$\ket{\rm v} \rightarrow - \ket{\rm v}$, 
where $\ket{\rm v}$ is the true ground state of the pp-wave theory,
whereas $\ket{0}$ is the string state corresponding
to the ground state for flat background (but not for $\m \neq 0$, 
where its energy is proportional to $\mu$).
In \cite{GKT}, the emergence of the previous parity 
assignment was explicitly shown in field theory. 

An important point should now be emphasised.
In order to use  and in particular to test \eqref{ham}
at the level of matrix elements,
it is essential  to construct the isomorphism
which connects the basis of states on which the string theory
Hamiltonian acts to the corresponding basis 
of field theory operators.
Lightcone string field theory is naturally equipped
with an orthonormal  basis of single- and multi-string states,
which does not  correspond to either the ``natural''
basis in field theory, i.e.~the basis of operators
with well-defined
conformal dimension (henceforth called the $\Delta$-BMN basis),
or to the basis originally considered by BMN.
This subtle  issue  required 
some time to be appreciated and fully understood
\cite{gross2,bits3,Gomis},
but  the isomorphism was finally constructed and used in
\cite{Gomis,Gomis2,GKT} to successfully test the correspondence.
Specifically, in \cite{GKT} the natural emergence of the
isomorphic to string basis was explained
by constructing the proper overlap of states in terms of
two-point function of BMN operators
where the conjugated BMN operator is defined through
hermitian conjugation {\it plus} an inversion
\cite{CKTvec}.
This procedure, which was  applied in \cite{GKT} to BMN operators
with scalar and/or vector impurities, will be 
used in section 3 in the case of fermion BMN operators.

Fermion BMN operators have  recently been studied, 
with an emphasis on mixing issues,
in \cite{Eden:2003sj,Bianchi:2003eg}, however 
the equivalence relation \eqref{ham}
has not been investigated so far in the fermionic part 
of the BMN sector of $\cN=4$  Yang-Mills.
The aim of this paper is to fill this important gap.
One of the main motivations for our analysis lies in the fact
that fermionic matrix elements
of $H_{\rm string}$ 
have never been compared to any field theory result.
Moreover, string field theory in the fermionic sector
is not just a simple extension of its bosonic
counterpart, and the construction of the
fermionic prefactor of the string field Hamiltonian is
not 
straightforward. Therefore, it is particularly compelling 
to investigate the BMN correspondence in the fermionic sector.
  
In this paper we will make use of the 
superstring vertex in the $SO(4) \times SO(4)$ formalism, 
whose construction was given in \cite{Panknew}.
It was shown in  \cite{Pankiewicz:2003ap}
that the $SO(8)$ formalism of \cite{SV1,SV2,AP,PS} 
is actually completely equivalent to 
the $SO(4) \times SO(4)$
construction, as it was already conjectured in
\cite{Panknew}.
The $SO(8)$ and the  $SO(4) \times SO(4)$ formalism
differ in that, in the former, the string interaction vertex 
is built upon the state  $\ket{0}$
(the ground state of the theory in flat background); 
whereas in the latter, 
the vertex is constructed on the true pp-wave ground state 
$\ket{\rm v}$. In both formalisms the external
string  states  are built on the true 
pp-wave ground state $\ket{\rm v}$.

Our analysis in field theory 
will be performed up to and including
$\cO (\l')$   in the
pp-wave 't Hooft coupling and
$\cO (g_2)$
in the genus counting parameter,
and hence incorporates string interactions
at the first nontrivial order.%
\footnote{This result, as well as  the results of 
\cite{Gomis,Gomis2,Gomis3,GKT} 
are in contradistinction with \cite{DiVecchia:2003yp},  
where it was argued that nonplanar effects (corresponding to 
string interactions in string theory) should not 
be incorporated into the pp-wave string theory/$\cN=4$ SYM 
duality.} 
Our result is simple: for all cases we consider,
the matrix elements of the string field theory Hamiltonian
derived from the superstring vertex of
\cite{SV1,SV2,PS,Panknew} agree
perfectly with the corresponding field theory quantities.
This result allows us to confirm the validity of the
conjectured duality relation
\eqref{ham} in the fermionic sector,  and  at the level 
of string interactions.

The plan of the rest of this paper is as follows.
In the next section we introduce and discuss 
BMN operators with two fermion impurities.
In section 3, we review in detail the procedure 
which leads to the identification of the basis in field theory 
which is isomorphic to the natural string basis 
of single- and multi-string states. We also summarise the strategy 
we follow in order to compare the matrix elements of 
the Yang-Mills dilatation operator to the matrix elements of the 
interacting string Hamiltonian --
with particular attention to the new features  
arising  from considering fermion BMN operators. 
In section 4 we derive the desired fermionic matrix elements 
of the string Hamiltonian in lightcone string field theory. 
Section 5 and 6 contain the calculation in field theory 
and the comparison to the string results previously derived.
We conclude with a few appendices  containing details 
of the calculations, as well as our 
notation and conventions 
in field and string  theory.

\section{Fermion BMN operators}
In order to study the BMN sector of $\cN=4$ Super Yang-Mills,
we need to pick an  R-charge subgroup $U(1)_J \subset SU(4)$.
Hence we need to decompose
$SU(4)\to
SO(4) \times U(1)_J \sim SU(2) \times SU(2) \times U(1)_J$.
The branching rule for the fermion representation ${\bf 4}$ of
$SU(4)$, to which the fermions 
$\l^{A}_{\a}$ of the $\cN=4$ theory belong, 
is \cite{Slansky:yr}
\beq
{\bf 4 } \longrightarrow {\bf (2,1)_{+} \ + \ (1,2)_{-} }
\ ,
\eeq
from  which we have, in terms of fields,
\beqa
\lambda^{A}_{\alpha} &\longrightarrow&
\big(\lambda_{r \alpha,\, (1/2)}\, , \,
\lambda_{\dot{r} \alpha, \, (-1/2)}
\bigr)
\ ,
\\
\cr
\bar{\lambda}_{A \dot{\alpha}}&\longrightarrow&
\big(\bar{\lambda}_{r \dot{\alpha},\, (-1/2)}\, , \,
\bar{\lambda}_{\dot{r} \dot{\alpha},\, (1/2)}\bigr)
\ .
\eeqa
Here $\alpha, \dot{\alpha} =1,2$ are spin indices,
$A=1\ldots 4$, and $r=3,4$, $\dot{r} = 1,2$.
Notice that there are four fermions with positive
R-charge,  $\lambda_{r \alpha,\, (1/2)}$  and
$\bar{\lambda}_{\dot{r} \dot{\alpha},\, (1/2)}$,
and four with negative R-charge,
$\lambda_{\dot{r} \alpha, \, (-1/2)}$,
$\bar{\lambda}_{r \dot{\alpha}, \, (-1/2)}$.
We will refer to the former
as to the BMN fermions, and to the latter as to the
anti-BMN fermions. To simplify the notation,
we will omit from now on the
$U(1)$ R-charge subscript in the fermion fields.

The following table  summarises the field content
(scalar, vector and fermion fields)
of the $\cN =4$ SYM theory participating in the BMN correspondence,
together with their canonical dimensions, R-charge and
decomposition into irreducible representations of
$SO(4) \times SO(4) \sim SU(2) \times SU(2)
\times SU(2) \times SU(2)$.
\begin{table}[ht]
\begin{center}
\begin{tabular}{|c|c|c|c|c|} \hline
{\sf field} & $\Delta_0$ & $J$ & $\Delta_0 - J$&
$SO(4)\times SO(4)$ \\ \hline
$Z$ & 1 & 1& 0 & $({\bf 1},{\bf 1})$ \\
$\bar{Z}$ & 1 & \!\!-1 & 2&  $({\bf 1},{\bf 1})$\\
$\phi^i$ & 1 & 0& 1& $({\bf 1},{\bf 4})$\\
$D_\mu$ & 1 & 0 & 1& $({\bf 4},{\bf 1})$\\
$\lambda_{r \alpha}$ & 3/2 & 1/2&1&
$\left(({\bf{2},1}),({\bf {2},1})\right)$ \\
$\bar{\lambda}_{\dot{r} \dot\alpha}$ & 3/2 & 1/2& 1&
$\left(({\bf 1,{2}}),({\bf 1,{2}})\right)$ \\
$\bar{\lambda}_{r \dot{\alpha}}
$ & 3/2 & \!\!-1/2& 2&
$\left(({\bf {2},1}) ,  ({\bf 1, {2}})\right)$ \\
$\lambda_{\dot{r} \alpha}$ & 3/2 & \!\!-1/2&2&
$\left(({\bf 1,{2}}) , ({\bf {2}, 1})\right)$ \\
\hline
\end{tabular}
\caption{
{\it In this table we list  the canonical dimension $\Delta_0$,
R-charge $J$
and $SO(4)\times SO(4)$ representations
for the fields of  $\cN=4$ Super Yang-Mills.
For convenience, we also write the corresponding
$\Delta_0 - J$ for each field.
}}
\label{table}
\end{center}
\end{table}

We now discuss the two-impurity fermion BMN operators.
Their precise form can be obtained by acting with two supersymmetry
transformations on the scalar BMN operators
\be
\label{7mar}
\cO_{ij , m}^J =\cC
\Bigl[ \sum_{l=0}^{J}e^{2\pi i ml \over J} \Tr \left(
\phi_i Z^l \phi_j Z^{J-l} \right)\ - \ \d_{ij} \,\Tr
(\bar{Z} Z^{J+1})
\Bigr] \ ,
\eeq
where $i,j=1,\ldots , 4$ and we have defined
\beq
\label{defofc}
\cC := {1\over  \sqrt{J N_0^{J+2}}}
\ , \qquad  N_0 := {\, g^2 \over  2}\,{N \over 4\pi^2} \ .
\eeq
The normalisation of the operators  is such that
their two-point functions take the canonical form
in the planar limit. This procedure
correctly identifies the possible compensating terms
which may be present in the expression of the operators.
We would like to remind the reader that these compensating terms
are crucial for a correct understanding of the dynamics
of the BMN sector. 

Specifically,
we will be considering
\beqa
\label{vacuum}
\cO_{\rm vac}^J &=& {1\over \sqrt{J N_{0}^{J}}}
\Tr Z^J \ ,
\\
\label{33}
\cO_{3 3;  m}^{ \alpha \beta; J}& =&\frac{\cC}{2}
\Bigl[ \sum_{l=0}^{J}e^{2\pi i ml \over J} \Tr (
\lambda_{3}^{ \alpha} \, Z^l
\, \lambda_{3}^{ \beta} \, Z^{J-l} )\Bigr] \ ,
\\
\label{34}
\cO_{3 4;  m}^{ \alpha \beta; J}& =&\frac{\cC}{2}
\Bigl[ \sum_{l=0}^{J}e^{2\pi i ml \over J} \Tr
( \lambda_{3}^{ \alpha} \, Z^l \, \lambda_{4}^{ \beta} \, Z^{J-l} )
\ - \
{\sqrt{2}\over 4} \Tr 
\Bigl( 
(F_{\mu \nu}\sigma^{\mu \nu} )_{\gamma}^{\, \beta}
\epsilon^{\alpha \gamma} \, Z^{J+1} \Bigr)
\Bigr]
\ .
 \eeqa
Very similar expressions can be written for operators
where $(3,4)\to (1,2)$, i.e.~undotted 
$SU(2)$ indices are replaced by dotted
ones.

In the following, we will also make extensive
use of the expressions for the double-trace operators
\beqa
\cT_{r \alpha, s \beta; \, m}^{J,y} &=& :
\cO_{r \alpha, s \beta; \, m}^{y \cdot J }:\
 : O_{\rm vac}^{(1-y)\cdot J}:
\ ,
\\ \cr
\cT_{\dot{r} \dot{\alpha}, \dot{s} \dot{\beta}; \, m}^{J,y}
&=&
:\cO_{\dot{r} \dot{\alpha}, \dot{s} \dot{\beta}; \, m}^{y \cdot J }:
\
: O_{\rm vac}^{(1-y)\cdot J}:
\ ,
\eeqa
where $y\in (0,1)$.

A few important comments are in order.
\begin{enumerate}
\item
First, note the appearance on right hand side of \eqref{34}
of an all-important compensating term which modifies the
na\"{\i}ve expression for $\cO_{34;  m}^{\alpha \beta; J}$.
Compensating terms are required in order
for the corresponding operators to
be conformal primaries in the BMN limit, and
are present also in the case of scalar
BMN operators \cite{Parnachev:2002kk}
(as the right hand side of
\eqref{7mar} for $i=j$ shows)
and vector BMN operators
\cite{gursoy,beisert}.
\item
Second, we would like to stress that
these compensating terms play a key r{\^o}le
in the evaluation of the conformal three-point functions
of vector  and mixed BMN operators.
Indeed, had they not been taken into account, one would
erroneously conclude that the three-point functions
for scalar, vector and mixed BMN operators 
take actually all the same form.
The three-point function coefficients for vector
and for mixed BMN operators were computed in \cite{CKTvec} and
\cite{GKT}, respectively, and found to be different from that
of the scalar case \cite{BKPSS}.%
\footnote{Their expressions are given in Eqs.~(3.29)-(3.31)
of \cite{GKT}.}
Of course, this is striking evidence against
a direct correspondence between the
conformal {\it three-point functions} and the superstring vertex
at the nontrivial, interacting level.
\item
Furthermore, the analysis of \cite{GKT} showed that
precisely thanks to the differences between
the three-point function coefficients for scalar, vector
and mixed impurity BMN operators it is possible
to reproduce,
from the field theory point of view, two key properties
of the three-string vertex of Spradlin and Volovich, namely:
\\
{\bf  a.} the vanishing of the three-string amplitude
for string states with one vector and one scalar impurity;
and \\
{\bf  b.} the relative minus sign in the string amplitude
involving states with two vector impurities
compared to that with two scalar impurities.\\
Once this is taken into account, perfect agreement
between the string and field theory predictions is found.
\end{enumerate}

To further clarify the r\^{o}le of the compensating terms,
we consider the flavour-singlet and flavour-triplet
combinations:
{\small
\beqa
\nonumber
\label{34S}
\!\!\!\!\!\!\!\!\!\!\!\!\!\!\!\!\!\!\!
\cO_{3 4, {\bf S};  m}^{\alpha \beta; J}& =&\frac{\cC}{2\sqrt{2}}
\Bigl[ \sum_{l=0}^{J}e^{2\pi i ml \over J} \Tr
( \lambda_{3}^{ \alpha} \, Z^l \, \lambda_{4}^{ \beta} \, Z^{J-l}
- 
\lambda_{4}^{ \alpha} \, Z^l \, \lambda_{3}^{ \beta} Z^{J-l}
)
\ - \
\frac{\sqrt{2}}{2} \Tr 
\Bigl(
(F_{\mu \nu}
\sigma^{\mu \nu} )_{\gamma}^{\, \beta}
\epsilon^{\alpha \gamma}
\, \, Z^{J+1}
\Bigr)
\Bigr]
\\ \\
\label{34T}
\!\!\!\!\!\!\!\!\!\!\!\!\!\!\!\!\!\!\!
\cO_{3 4, {\bf T};  m}^{ \alpha \beta; J}& =&\frac{\cC}{2\sqrt{2}}
\Bigl[ \sum_{l=0}^{J}e^{2\pi i ml \over J} \Tr
( \lambda_{3}^{ \alpha} \, Z^l \, \lambda_{4}^{ \beta} \, Z^{J-l}
\, + \,
\lambda_{4}^{ \alpha} \, Z^l \, \lambda_{3}^{ \beta} \, Z^{J-l}
)
\Bigr]
\ .
\eeqa
}
We can further decompose each of the
two operators in
\eqref{34S} and \eqref{34T} into
singlet and triplet of the spin, that is
\beqa
\label{decS}
\cO_{3 4, {\bf S};  m}^{\alpha \beta; J} &\longrightarrow&
{\bf (1, 3^{+}) \ +\ (1,1)} \ ,
\\
\label{decT}
 \cO_{3 4, {\bf T};  m}^{\alpha \beta; J}
&\longrightarrow&
{\bf (3^{+}, 3^{+}) \ +\ (3^{+},1)} \ .
\eeqa
It is immediately seen that the compensating term
on the right hand side of \eqref{34S} is
symmetric under the exchange of the spin indices $\alpha$
and $\beta$. This means that this compensating term will affect
only the ${\bf (1, 3^+ )}$ representation.
This is perfectly consistent with the decomposition
of the two-impurity BMN operators with {\it vector} impurities
according to irreducible representations of $SO(4) \times SO(4)$.
Indeed, by combining two vector  impurities we can
form the following representations:
\beq
\label{decvec}
{\bf
(1,4) \times (1,4) \ = \ (1,1) \ +\ (1,9) \ + \
(1, 3^+ ) \ + \ (1, 3^- ) } \ .
\eeq
The only representation the right hand sides of
\eqref{decS} and \eqref{decT}
have in common with the right hand side of \eqref{decvec} are
precisely ${\bf (1, 3^+ )} $, which
receives a compensating term, and ${\bf (1,1)}$,
for which however no compensating term is generated
as this order.%
\footnote{For a discussion along similar lines on
the possible mixing of fermion BMN operators
with scalar operators, see \cite{Bianchi:2003eg}.}

For completeness, we mention here what are the possible
irreducible representations of
$SO(4) \times SO(4)\sim SU(2) \times SU(2)\times SU(2) \times SU(2)$
that can be obtained by combining two
fermion impurities:%
\footnote{See also the discussion in section IV of
\cite{Shahin}.}
\beqa\label{decomp1}
{\bf ((2,1), (2,1)) \times ((2,1), (2,1)) } &=&
{\bf (1,1) \ + (3^+ , 3^+) \ + \ (3^+ , 1) \ +\ (1 , 3^+)} \ ,
\\\label{decomp2}
{\bf
((1,2), (1,2)) \times ((1,2), (1,2)) } &=&
{\bf (1,1) \ + (3^- , 3^-) \ + \ (3^- , 1) \ +\ (1 , 3^- )} \ ,
\\\label{decomp3}
{\bf
((2,1),(2,1)) \times ((1,2), (1,2))}&=&
{\bf (4,4)} \ .
\eeqa

\section{Comparing matrix elements of  $H_{\rm string}$ and 
$\Delta$}
In this section we briefly review the
strategy adopted in \cite{Gomis,GKT}  to compare matrix elements
of the dilatation operator $\Delta$ in SYM to  matrix elements of
the fully interacting string Hamiltonian  $H_{\rm string}$. 

To begin with, we notice that  ${H}_{\rm string}$
and $\Delta$ act on the states of two different theories.
Hence, in order to test the duality \eqref{ham} we need
to construct an isomorphism between the Hilbert spaces of
the lightcone pp-wave string field theory
and of the BMN sector of  $\cN=4$ Yang-Mills.
The choice of the basis in string theory and in
field theory is in principle arbitrary, however we remember that
string field theory is naturally equipped with an orthonormal  basis
of single- and 
multi-string states, given by   the tensor product
of single-string states.
This natural string basis, $\{ \ket{s_\alpha}^{\rm string} \}$,  
diagonalises the free string Hamiltonian. 
Once interactions are taken into account, 
they allow the strings to split and join,
i.e.~the states in the natural string basis are not eigenstates
of the  interacting pp-wave string Hamiltonian $H_{\rm string}$.
Matrix elements of $H_{\rm string}$ 
are known in the natural string basis,
hence our  goal will be 
the identification of the field theory basis
$\{ \ket{s_\alpha}^{\rm SYM} \}$
which is isomorphic to it.
This, in turns, will enable us to recast \eqref{ham} 
at the level of the matrix elements as 
\be 
\label{ham2}
{}^{\rm string}\bra{s_\alpha}\mu^{-1}{H}_{\rm string}
\ket{s_\beta}^{\rm string}\, =\,
{}^{\rm SYM}\bra{s_\alpha} \Delta - J \ket{s_\beta}^{\rm SYM}\ .
\ee

But what is the situation in field theory?  In the conformal
$\cN=4$ Yang-Mills there is also a privileged basis of states,
the basis of conformal primary BMN operators $\cO_{\D_\alpha}(x)$,
or  $\Delta$-BMN basis. This is the basis of  the eigenstates of
the SYM dilatation operator, and its eigenkets can be expressed as
linear combinations  of the original BMN operators $\cO_\alpha
(x)$ proposed in \cite{BMN}. For BMN operators with scalar
impurities, the  $\Delta$-BMN basis was explicitly constructed in
\cite{BKPSS}, and extended to include vector and mixed impurities
in \cite{CKTvec}. Conformal invariance guarantees that,  in the
$\Delta$-BMN basis, two- and three-point functions of $\Delta$-BMN
operators take the canonical form, with a universal
$x$-dependence.  Specifically, for scalar conformal primary
operators one has
\begin{equation}
\label{2pt} 
\langle {\cO}^{\dagger}_{\D_\alpha} (x) \cO_{\D_\beta}
(0) \rangle = \frac{\d_{\alpha \beta}}{(x^2)^{\Delta_\alpha}} \ ,
\end{equation}
\begin{equation}
\label{3pt} \langle \cO_{\Delta_1}(x_1) \cO_{\Delta_2}(x_2)
\cO^{\dagger}_{\Delta_3}(x_3) \rangle  = \frac{C_{123} }
{(x_{12}^2)^{\frac{\D_1+\D_2 -\D_3}{2}}
 (x_{13}^2)^{\frac{\D_1+\D_3 -\D_2}{2}}
 (x_{23}^2)^{\frac{\D_2+\D_3 -\D_1}{2}}} \ .
\end{equation}
Correlation functions of conformal primary operators with vector
or mixed  impurities appear to be 
harder to interpret, however it was noted in \cite{CKTvec} that
this problem  is eliminated, and the correlators for all types of
impurities can be expressed in a form similar to \eqref{2pt} and
\eqref{3pt}, if on the left hand sides of \eqref{2pt} and
\eqref{3pt} we use a different notion of conjugation $\bar{\cO}$
instead of $\cO^\dagger$ \cite{CKTvec,DiVecchia:2003yp}. 
This  different  notion of
operator conjugation is defined as {\it hermitian conjugation}
followed by an  {\it inversion} of the insertion point of the
operator $x'_{\mu} = x_{\mu} / x^2$.

Let us now briefly review the transformation properties under an
inversion of scalar and fermion operators \cite{Fradkin:is}. 
A scalar operator
$\cO_{\Delta}(x)$ of conformal dimension $\Delta$ transforms as
\beq
\label{ccc}
 \cO_{\Delta} (x) \rightarrow
\cO_{\Delta}^{'} (x') = x^{2 \Delta}
\cO_{\Delta}(x) \ \ , \ \ x_{\m}\to x'_{\m} = {x_{\m}
\over x^2} \ ,
\eeq
while  vector or tensor operators
(i.e.~operator with vector impurities) pick a factor
$J_{\mu\nu}(x) =\d_{\mu\nu}-2x_{\mu}x_{\nu}/x^2$ on the right hand
side for each vector index of the operator.  $J_{\mu\nu}(x)$ is
the usual inversion tensor, in terms of which the Jacobian of the
inversion is expressed $\partial x'_{\mu}  / \partial x_{\nu} =
J_{\mu \nu} (x) / x^2 $. This prescription is essential in order
to make vector $\Delta$-BMN operators orthonormalisable, see
section 2 of \cite{CKTvec} for more details. 
Let us now concentrate on  the fermionic
operators  which are of direct concern for 
this paper.%
\footnote{The transformation under  inversion of 
fermionic BMN operators  has  also been considered in 
\cite{DiVecchia:2003yp}.}
It is well known that, under conformal inversion, 
a Dirac spinor field $\psi$ of dimension $d$ 
transforms as   \cite{Fradkin:is}
\beq
\label{Dirac}
\psi (x) \rightarrow \psi' (x') = 
\eta \, {\hat{x}\over |x|} \, x^{2d} \, \psi (x)
\ , \qquad \eta^4 =1 \ , 
\eeq
where $\hat{x}=  x_\m \gamma^\m$, and $\gamma^\m$ 
are the Euclidean gamma matrices. In terms 
of Weyl spinors $\lambda_\a$, $\bar{\xi}^{\dot\alpha}$, 
\eqref{Dirac} implies
\beqa
\label{Weyl}
\l_{\a} (x)  \rightarrow \l_{\a}' (x') &=& 
\eta \, {(x \bar{\xi})_{\a}\over |x|}   \, x^{2d}
\ , \qquad 
\\
\bar{\xi}^{\dot{\alpha}} (x) \rightarrow \bar{\xi}^{\dot\a '} (x')
&=&
\eta \, {(\bar{x} \l)^{\dot\a}\over |x|}  
\, x^{2d}
\ ,
\eeqa
where we set $x= x_\m \s^{\m}$, $\bar{x} = x_\m \bar{\s}^\m$.
Hence, an operator of conformal dimension $\Delta$ 
with  $f=p+q$ fermionic insertions transforms under inversion as:
\beq
\cO_{\a_1 \ldots \a_{p}}^{\dot{\a}_1 \ldots  \dot{\a}_{q}}
 (x)  \, \rightarrow \,  
\eta^f  
\, 
x^{2 \Delta} \, 
{J_{\a_1 \dot{\b}_1}} \cdots
{J_{\a_p \dot{\b}_p}}\cdots
{\bar{J}^{\dot{\a}_1 {\b}_1}}\cdots
{\bar{J}^{\dot{\a}_q {\b}_q}}
\,  
\cO_{\b_1 \ldots \b_{q}}^{\dot{\b}_1 \ldots  \dot{\b}_{p}}
 (x)
\ ,
\eeq
where
\beq
J_{\a\dot{\b}}(x)\,  := \,
{x_{\a \dot\b} \over |x|}   \ , \qquad
\bar{J}^{\dot{\a} {\b}}(x) \, := 
\, 
{\bar{x}^{\dot\a \b} \over |x|}
\ .
\eeq
The notion of conjugation  \cite{CKTvec},  
as ordinary hermitian conjugation 
followed by an inversion,  can then applied to a generic 
operator with conformal dimension $\Delta$
with scalar, vector or fermion impurities; and  
the conjugated operator $\bar{\cO}$
can  then be 
written (schematically) as:  
\beq 
\label{scalinvnewww} 
\bar\cO_{\Delta} (x) \,
\equiv \, x^{2 \Delta } \, J \cdot \cO^\dagger_{\Delta}(x)
\ , 
\eeq 
where by $J$ we mean the tensor product by 
the appropriate inversion operators
$J_{\mu\nu}(x)$,  for each vector index, 
and $J_{\a \dot\a} (x)$ (or  $\bar{J}^{\dot\a \a}(x)$), 
for  each spinor index.
It was noticed in \cite{CKTvec} that 
the advantage of this new conjugation   
resides in that 
the two-point function for
scalar, vector and fermion $\Delta$-BMN operators take all the
same canonical  form when the $\bar{\cO}$ operator is employed:
\begin{equation}
\label{2ptbar}
\langle \bar{\cO}_{\D_a} (x) \cO_{\D_b}
(0) \rangle \ = \  {\d_{a b}} 
\ .
\end{equation}
The right hand side of \eqref{2ptbar}   does not
depend on $x$,  and represents the overlap of the corresponding
states in conformal $\cN=4$ Super  Yang-Mills.%
\footnote{A side comment: for two-impurity fermion 
BMN operators, 
an additional minus sign should be included in the definition of
the hermitian conjugation of the operators in order to get 
the two-point functions normalised as in \eqref{2ptbar}.}

The operators of a generic basis in field theory will not enjoy
the simple interpretation in terms of states given by
\eqref{2ptbar}, nevertheless we can always 
decompose these operators along
the eigenvectors of the $\Delta$-BMN basis. Specifically, 
for any operator basis $\tilde\cO_\alpha$ such that 
\be
\label{16} 
\tilde\cO_\alpha \ = \ 
U_{\alpha \beta}\,\cO_{\D_\beta}
\ , 
\ee 
where $U_{\alpha \beta}$ is a constant matrix,
we can   easily compute  the
overlap of the basis vectors: this   
is simply given by \cite{GKT} 
\be
\label{smths} 
\langle \bar{\tilde\cO}_\alpha (x) \tilde\cO_\beta
(0) \rangle \ = \
 U_{\beta \gamma} U^\dagger_{\gamma \alpha} \ \equiv \
 S_{\beta \alpha} \ .
\ee
The previous relation is highly nontrivial: despite the fact that
the operators $\tilde\cO_\alpha$ do not have definite scaling
dimensions, the overlap \eqref{smths} does not depend on $x$! 
The key step in this procedure is to be found in the inversion
procedure. Indeed, notice that  in \eqref{ccc} the {\it full}
conformal dimension $\Delta$ of the conformal BMN operator
$\cO_{\Delta_\a}$ is used in order to define the proper notion of
conjugation. This expression will have in general a perturbative
(and, in principle, nonperturbative) expansion in $\lambda'$,
which is eventually responsible for the simple form of
\eqref{smths} (and \eqref{2ptbar}).

A  simple and practical way of calculating simultaneously the
overlaps of states and the matrix elements of the anomalous
dimension operator $\delta=\Delta-\Delta_{\rm cl}$ (where
$\Delta_{\rm cl}$ is the  canonical dimension in the free theory)
was described in \cite{GKT}. We will apply this procedure in the
following, therefore we briefly outline its steps. First, one
defines the barred-operator $\bar{\tilde\cO} (x)$ as the Hermitian
conjugation of $\tilde\cO (x)$ followed by an inversion of the
resulting operator, defined as if it was free, i.e.~instead of the
factor $x^{2 \Delta}$ in \eqref{scalinvnewww} 
we put $x^{2 \Delta_{\rm cl}}$, such that 
\beq 
\label{scalinvnew} 
\bar\cO_{\Delta} (x) \,
\equiv \, x^{2 \Delta_{\rm cl}} \, J \cdot \cO^\dagger_{\Delta}(x)
\ .
\eeq 
The two-point function takes now the form:
\beqa
\label{genres} 
\langle
\bar{\tilde\cO}_\alpha (x) \tilde\cO_\beta (0) \rangle & = &
 U_{\beta \gamma}\, e^{\delta_\gamma \log (\Lambda x)^{-2}}\,
U^\dagger_{\gamma \alpha} 
\nonumber \\
&=&
S_{\beta\alpha} + T_{\beta\alpha} \log (\Lambda x)^{-2} +
\cO \,
\bigl((\log (\Lambda x)^{-2})^2\bigr
)\ ,
 \eeqa
where we have expanded the full result in powers of $\log x^{-2}$.
{}From \eqref{genres} we
can read off the overlap of the two states, defined as the
zeroth-order term of the expansion, 
\beq 
\label{esse}
S_{\beta\alpha}=U_{\beta
\gamma} U^\dagger_{\gamma \alpha} \ , 
\eeq 
as well as  the matrix
of anomalous dimensions in this basis,  given by the first order
term, 
\beq 
\label{ti}
T_{\beta\alpha}=U_{\beta \gamma} \delta_\gamma
U^\dagger_{\gamma \alpha} \ . 
\eeq 
We now consider the original
BMN basis, for which we have 
\be 
\label{origres} 
\langle
\bar{\cO}_\alpha (x) \, \cO_\beta \rangle \ = \
 S_{\beta\alpha} + T_{\beta\alpha} \log x^{-2} +
 \cdots \ ,
\ee
and relate this basis to the isomorphic to string basis via
a transformation $U$ as in \eqref{16}: 
\beq 
\cO^{\rm string}_\beta= U_{\beta \gamma}
\cO_\gamma \ \ , \qquad \bar{\cO}^{\rm string}_\alpha =
\bar{\cO}_\delta U^\dagger_{\delta \alpha} \ . 
\ee 
In the basis
isomorphic to the natural string basis, we have
\be 
S^{\rm string}
= \uno = U S U^\dagger \ , \qquad T^{\rm string} =  U T U^\dagger
\ . 
\eeq 
Notice that $S$ is a  Hermitian, positive matrix,
therefore $S^{-1/2}$ is well-defined.%
\footnote{The matrix $ S^{-{1\over 2}}$ also appears in
\cite{bits3} and \cite{SV3}.} $S$ is  then diagonalised by
choosing 
\beq
U\, := \, S^{-1/2} \cdot V \ , 
\eeq
where $V^\dagger V=\uno$: 
\beqa 
S
&\longrightarrow& USU^\dagger \, = \, \uno \ \ ,
\\
T &\longrightarrow& UTU^\dagger \, = \, 
V^\dagger (S^{-{1\over 2}} T
S^{-{1\over 2}})V \ \ . 
\eeqa 
At this point, there is still an  arbitrariness contained in $V$. 
This was fixed in \cite{gross2,Gomis} by requiring that 
\eqref{ham2}
holds and that the known three-string interaction vertex of the
pp-wave light-cone string field theory \cite{SV1,SV2} is
reproduced from gauge theory matrix elements involving BMN states
(operators) with two scalar impurities. This condition implies
$V=\uno$. 
Hence, we conclude that 
the matrix of anomalous dimensions in the string
basis is given by \cite{Gomis,GKT}
\beq 
\label{Gamma} 
\Gamma := T^{\rm string} =
S^{-{1\over 2}} \, T \, S^{-{1\over 2}} \ . 
\eeq 
This is the result we wanted to derive. 
The procedure for obtaining  the matrix elements 
of the anomalous dimension operator (and hence 
of the dilatation operator in Yang-Mills) 
in the basis in field theory which is isomorphic to 
the natural string basis is therefore clear: 
choose an arbitrary operator basis in Yang-Mills
(in particular the simple basis originally considered 
by BMN in \cite{BMN}), and
compute two-point functions as in \eqref{genres} 
(or \eqref{origres}); 
work out the expressions for the matrices $S$ and $T$; 
and, finally,  
apply \eqref{Gamma} to derive the desired matrix elements
in the isomorphic to string basis.

Eq.~\eqref{Gamma} was employed in
\cite{GKT} to explore and successfully test the bosonic (scalar
and vector) sector of the BMN correspondence, finding precise
agreement between the matrix elements in of the interacting string
Hamiltonian in string field theory and the dual matrix elements
\eqref{Gamma} of the field theory dilatation operator. 
In the next sections we will pursue this programme 
and carry out a number of
tests in the fermionic sector of the string and field theory. 
More precisely, we will show that, with the same choice of 
$V=\uno$, the matrix elements of $\Gamma$
between BMN operators with 
two fermion impurities 
precisely agree with the corresponding matrix elements of the
interacting string Hamiltonian. 

Other studies of the dilatation operator in $\cN=4$ Yang-Mills 
and its interpretation in quantum mechanical and integrable 
models can be found in the recent papers
\cite{Janik,Minahan:2002ve,B1,SV3,B2,Beisert:2003jj,
Beisert:2003yb,Beisert:2003ys,Kristjansen:2004ei, Arutyunov:2004xy}.

\section{Matrix elements of $H_{\rm string}$ in string field theory}
There are two equivalent ways  to describe
superstring interactions in string field theory,
known as the $SO(8)$ and the $SO(4) \times SO(4)$ formalism.
In the former approach,
the  three-string vertex in string field theory
is built upon a state $\ket{0}$ with energy equal to $4 \mu$.
This state is therefore not the ground state,
but it has the advantage that,  as $\mu \to 0$,
the $SO(8)$ construction flows smoothly
to   string field theory in flat space
\cite{Mandelstam:hk,Green:1982tc,Green:hw,Green:fu}.
In the  $SO(4) \times SO(4)$ formalism, the Hilbert space of states
in string field theory is built on the true vacuum
$\ket{\rv}$ of
pp-wave string theory (see Appendix A for details).
Remarkably, the two formalisms have been shown to
be completely equivalent \cite{Panknew}, hence it is only
a matter of convenience which one to use.
In this paper we will make use of the $SO(4) \times SO(4)$
vertex,  since there it is more straightforward to
compute string amplitudes involving fermionic oscillators.

The string amplitude has the form \cite{SV1,SV2}%
\footnote{The reader is referred to
Appendix A for more details on the string field theory vertex.}
\bea
\bra{1} \bra{2} \, H_{\rm string}\,  \ket{3}
\ = \
\langle\Phi| {\sf P} \ket{V_B}\ket{V_F} \ ,
\eea
where $\langle\Phi|:=
\langle 1|\langle 2|\langle 3|^{'}$
is the external three--string state,
and $\ket{V_B}$ and $\ket{V_F}$
are  the kinematical  part of the bosonic and fermionic
vertex,  \eqref{Vb} and  \eqref{Vf} respectively.
Finally, the prefactor {\sf P}  is written in
\eqref{pref}.

We will be interested in external states
with fermionic impurities,
\beq
\beta^{\alpha \beta \, \dagger}_{n (\rr)}
\, \beta^{\alpha' \beta' \, \dagger}_{-n (\rr)}
\, 
\ket{\rv}_{\rr}
\ , 
\eeq 
where the fermionic operators $\beta$'s
are related to the oscillators
in the string basis by \cite{PS}
\beq
\beta_{n} \, = \, \frac{1}{\sqrt{2}} ( b_n + i b_{-n}) \ ,
\ \ \ \
\beta_{-n} \, = \, \frac{1}{\sqrt{2}} ( b_n - i b_{-n}) \ .
\eeq
Specifically, we will compute matrix elements of the form
\beq
{\cal H}^{r \a , s \b ;\,  m, J}_{r' \a' , s' \b' ; n, J, y}
\ := \  
{1\over \mu} \langle\cT_{r'\alpha', s' \beta'; \,  n\,}^{J,y}
| H_{\rm string}  |
\cO^{r \alpha, s \beta;  m}_{J}\rangle
\ ,
\eeq
for all possible values of the indices. 
After a lengthy but straightforward algebra, we obtain:
\bea 
\label{sftr}
&&
\langle {\rm v} 
|\b_{\a\b,-m(3)}\b_{\gamma\d,m(3)}
\b_{\a'\b',n(1)}\b_{\gamma'\d',-n(1)}|H_3\rangle
\ = \ C_{\rm norm}\frac{\b+1}{3\pi^2 \mu}\sin^2{\pi m \b}
\\
&&\cdot
\Bigl[ \e_{\a'\a}\e_{\gamma'\gamma}
(\e_{\d\b}\e_{\d'\b'}+\e_{\d\b'}\e_{\d'\b})
+
\e_{\gamma'\a}\e_{\a'\gamma}(\e_{\d\b}\e_{\d'\b'}+\e_{\d\d'}\e_{\b\b'})
- \e_{\a\gamma}\e_{\a'\gamma'}(\e_{\b\b'}
\e_{\d\d'}-\e_{\d\b'}\e_{\d'\b})\Bigr]
\, , 
\nonumber
\eea
where $C_{\rm norm}$ is given in \eqref{ccp}. 
An expression  similar to \eqref{sftr}  holds when 
the fermions in \eqref{sftr} have both 
dotted indices. Notice that from \eqref{sftr} it follows  that 
\\
{\bf a.} 
the string amplitude vanishes 
whenever a fermion appears more than once, whereas 
\\
{\bf b.} 
it is nonvanishing when all fermions are different, and gives 
always the same  result (up to a minus sign). 

It is more illuminating to write \eqref{sftr} for
a few basic cases:
\bea
\label{res1}
{\cal H}^{1 2 , 11 ;\,  m, J}_{11 , 12 ; n, J, y}
&:=& 
\m^{-1}
\langle {\rm v}|\b_{-m(3)}^{12}\b_{m(3)}^{11}
\b_{11,n(1)}\b_{12,-n(1)}
|H_3\rangle
\\ 
\nonumber
&=&
-\m^{-1}
\langle {\rm v}|\b_{21,-m(3)}\b_{22,m(3)}\b_{11,n(1)}\b_{12,-n(1)}
|H_3\rangle
\ = \ 
- \,\l' \, 
 C_{\rm norm}\frac{\b+1}{\pi^2 }\sin^2{\pi m \b} \ , 
\\ \cr
\label{res2}
{\cal H}^{11, 11 ;\,  m, J}_{11 , 11 ; n, J, y}
&:=&
\nonumber 
\m^{-1}
\langle {\rm v}|\b_{-m(3)}^{11}\b_{m(3)}^{11}
\b_{11,n(1)}\b_{11,-n(1)}|H_3\rangle 
\nonumber \\
& =& 
\m^{-1}
\langle {\rm v}|\b_{22,-m(3)}\b_{22,m(3)}\b_{11,n(1)}\b_{11,-n(1)}
|H_3\rangle \ = \ 0 \ , 
\\ \cr
\label{res3}
{\cal H}^{22, 11 ;\,  m, J}_{11 , 22 ; n, J, y}
&:=&
\m^{-1}
\langle {\rm v}|\b_{-m(3)}^{22}\b_{m(3)}^{11}\b_{11,n(1)}
\b_{22,-n(1)}|H_3\rangle 
\\ \nonumber 
&=& 
\m^{-1}
\langle {\rm v}|\b_{11,-m(3)}\b_{22,m(3)}\b_{11,n(1)}
\b_{22,-n(1)}|H_3\rangle 
\ = \  0 \ , 
\\ \cr
\label{res4}
{\cal H}^{21, 11 ;\,  m, J}_{11 , 21 ; n, J, y}
&:=&
\m^{-1}
\langle {\rm v}|\b_{-m(3)}^{21}\b_{m(3)}^{11}
\b_{11,n(1)}\b_{21,-n(1)}|H_3\rangle 
\\ \nonumber 
&=&
-\m^{-1}
\langle {\rm v}|\b_{12,-m(3)}\b_{22,m(3)}\b_{11,n(1)}\b_{21,-n(1)}
|H_3\rangle
\ = \ 
\l' \, C_{\rm norm}\frac{\b+1}{\pi^2 }\sin^2{\pi m \b}
\ . 
\eea
We can directly compare 
the expressions \eqref{res1}-\eqref{res4} 
to the analogous matrix elements obtained
from the three-string vertex of \cite{SV1,SV2}
for scalar 
and for vector and mixed 
(scalar-vector) BMN states: 
\beqa
\label{sv1}
{1\over \mu}
\langle
\cT_{ij, n}^{J,y}  | H_{\rm string}  | \cO_{ij, m}^{J}\rangle
&= & -\, \l' \, 
C_{\rm norm} \frac{\b+1}{\pi^2 }\sin^2{\pi m \b}
\ ,
\\ \cr
\label{sv2}
{1\over \mu}
\langle
\cT_{i \m, n}^{J,y} | H_{\rm string} | \cO_{i \m, m}^{J}\rangle
& = &  0
\ ,
\\ \cr
\label{sv3}
{1\over \mu}
\langle
\cT_{\m \n , n}^{J,y} | H_{\rm string} | \cO_{\m \n, m}^{J}\rangle
& = & \l' \, C_{\rm norm}  \frac{\b+1}{\pi^2 }\sin^2{\pi m \b}
\ .
\eeqa
Notice hat the amplitude in \eqref{res1} is
identical  to the amplitude \eqref{sv1} involving 
two scalars of different flavour, 
whereas that in \eqref{res4} is equal to the amplitude
\eqref{sv3}
involving two different vectors. Similar considerations apply to 
the vanishing of the mixed amplitude \eqref{sv2} and 
\eqref{res2}, \eqref{res3}.

The equality of the string amplitude between two BMN states with 
scalar impurities and two BMN states with fermion impurities
had already been derived, in the $SO(8)$ formalism, 
in \cite{PS}. We also notice that our amplitudes, derived in the 
$SO(4) \times SO(4)$ formalism, precisely coincide with 
those of \cite{PS}.

\section{Matrix elements of $\Delta$ in $\cN=4$ Yang-Mills}
In \cite{GKT}, a general technique was devised for deriving 
the matrix of overlaps $S$ \eqref{esse}, 
the matrix of anomalous dimension $T$ \eqref{ti}, 
and hence  the desired 
matrix elements of the SYM dilatation operator in
the isomorphic to string basis, \eqref{Gamma}, from the 
coefficients of the 
{\it three-}point functions of BMN operators.
Here we report the results of the analysis of \cite{GKT}, 
referring the curious reader to section 3 of that paper
for more details.%
\footnote{Our notation and conventions in Yang-Mills 
are reviewed in Appendix A.}

The matrices $S$ and $T$ have an expansion in powers of $g_2$, 
but in our analysis we will only 
need their expressions up to and including 
$\cO (g_2 )$ terms. We will also work at one loop 
in the effective 't Hooft coupling $\l'$. Notice that  
the matrix $T$ is of $\cO (\l' )$, whereas $S$ is of $\cO (1 )$. 
In this case, \eqref{genres} is simply
\beq
\label{stdef}
\langle \cO_{\a}(0) \bar{\cO}_\b (x)\rangle = 
 S_{\a \b} + T_{\a \b} \log (x \Lambda)^{-2}
\ . 
\eeq
Let us focus on the following  
{\it three-}point correlators, 
\beq
\label{three}
G (x_1,  x_2,  x_3 ) = 
\langle 
\cO_{AB, n}^{y\cdot J} (x_1) 
\cO_{\rm vac}^{(1-y)\cdot J}(x_2) 
 \bar{\cO}_{AB, m}^J (x_3)
\rangle \ ,  
\eeq
where $A$ can be a scalar, vector or fermion index, 
and $A\neq B$.  On general grounds, 
these three-point function take  the form 
\cite{Constable1,BKPSS,CKT,CKT2,Constable2}
\beq 
\label{42}
G (x_1,  x_2,  x_3 ) =  g_2 C_{m, n y}
\left[ 1 - \l'\left( a_{m,ny} \log (x_{31} \Lambda)^2 + 
b_{m,ny} \log (x_{32} x_{31} \Lambda / x_{12}) \right) \right]
\ , 
\eeq
where $g_2 C_{m,ny}$ is the tree-level contribution, with
\beq
\label{def-C-tree}
C_{m,ny}:=
{  \sqrt{(1-y)/y}\,  \sin^2 (\pi m y) \over
\sqrt{J} \, \pi^2 (m-n / y)^2}
\ , 
\eeq
and 
the coefficients $a_{m,ny}$ and $b_{m,ny} $ must be 
calculated in perturbation theory at $\cO (\l')$.
The {\it two-}point function
$\langle \cT_{AB, n}^{J,y} (0) \, \bar{\cO}_{AB, m}^J (x) \rangle$ 
can easily be derived from \eqref{three} by simply setting 
$x_{13}=x_{23} = x$ and $x_{12}= \Lambda^{-1}$ 
\cite{Constable2}, 
\beq
\label{a+b}
\langle \cT_{AB, n}^{J,y} (0)\,  \bar{\cO}_{AB, m}^J (x) \rangle
= g_2 C_{m,ny}
\left[ 1 + \l'\left( a_{m,ny} +b_{m,ny} \right)\log (x \Lambda)^{-2}  
\right]
\ .
\eeq
The analysis of section 3 of \cite{GKT} then showed that 
the matrices $S$ and $T$ are then given, up to $\cO (g_2)$, 
by the following expressions:
\beqa
\label{matrixS}
S &=& 
\left(\begin{tabular}{cc}
$\d_{mn}$ & $g_2\,  C_{m,qz}$  \cr \cr
$ g_2 \, C_{py,n}$ &  $\d_{pq}$
\end{tabular}\right) \ + \ \cO (g_2^2) \ = \ \uno + g_2 s +\cO (g_2^2)  \ , 
\\ \\
\label{matrixT}
T &=& \l' \left(\begin{tabular}{cc}
$m^2\, \d_{mn}$ & $g_2\,  C_{m,ny} \, ( a + b)_{m,qz}$  \cr \cr
$g_2\,  C_{py,n}\, (a+b)_{py,n}$  & $ (p^2 / y^2)\, \d_{pq} \d_{yz}$
\end{tabular}\right) \ + \ \cO (g_2^2)  
\\ \nonumber \cr
&\equiv & d\  + \ g_2 t  \ + \ \cO (g_2^2) 
\ , 
\eeqa
with 
\beqa
\label{matrix-d}
d &=& \l' \left(\begin{tabular}{cc}
$m^2\, \d_{mn}$ & $0 $  \cr \cr
$0 $  & $ (p^2 / y^2)\, \d_{pq} \d_{yz}$ 
\end{tabular}\right) \ , 
\\   
\cr
\label{matrix-t}
t &=& \l' \left(\begin{tabular}{cc}
$0 $ & $C_{m,ny} \, ( a + b)_{m,qz}$  \cr \cr
$ C_{py,n}\, (a+b)_{py,n}$  & $0$
\end{tabular}\right)   
\ .
\eeqa
It then follows that  
\beq
S^{-1/2} = \uno - g_2 (s/ 2) + \cO (g_2^2)
\eeq
diagonalises $S$ at  $\cO (g_2)$, and hence
\eqref{Gamma} leads to 
\beq
\label{Gammap}
\Gamma \ =\ 
d + g_2 \Bigl[ t  -{1\over 2} \{s \, , \, d \}
\Bigr]
\ . 
\eeq

We now need to compute the explicit expressions for 
$a_{mn}^y$ and $b_{mn}^y$ for the fermion case.
First, it was observed in \cite{GKT} that,  
at  $\cO (\l')$ in planar perturbation theory,   
the coefficient $a_{mn}^y$ is simply given by 
 the $\cO (\l')$ anomalous dimension
of the ``small'' BMN operator at $x_1$. 
In accordance with supersymmetry, it turns out that
the $\cO (\l')$ anomalous dimension of BMN operators with 
two arbitrary impurities is given by
\beq
\label{a-all}
a_{m,ny}  
\ = \  {n^2 \over y^2} \ , 
\eeq 
independently of the type of impurity considered. 
This result was first obtained for the case of two 
scalar impurities in \cite{BMN}, for one scalar and one vector
impurity in \cite{gursoy}, for two vector impurities
in \cite{klose}. We have also explicitly derived \eqref{a-all} 
in the fermion case with a perturbative calculation in Yang-Mills.

We now move on to consider $ b_{m,ny}$. 
{}From \eqref{42}, we see that 
$ b_{m,ny} $ is the coefficient which multiplies 
the $\log x_{12}$ contribution in the three-point function
$G(x_1 , x_2 , x_3)$ in \eqref{three}, 
where the ``large'' (``small'') 
BMN operator $\bar{\cO}_{AB, m}^J$
($\cO_{AB, n}^{y\cdot J} $)
is inserted at $x_3$ ($x_1$), and the vacuum operator
$\cO_{\rm vac}$ at point $x_2$. 
Hence, the tactic we will follow in the next section will 
consist in computing the $\log x_{12}$ term of 
the three-point function%
\footnote{It was shown in \cite{GK,GKT} that from the knowledge 
of the coefficient $ b_{m,ny}$ it is also possible to determine 
the coefficient of the {\it conformal} three-point function, despite the 
fact that, due to mixing effects \cite{Bianchi,BKPSS}, 
the correlator \eqref{three} does not take the conformal form
\eqref{3pt}, since the original BMN operators in \eqref{42}
are not conformal primaries for $g_2 \neq 0$.}
$G(x_1, x_2, x_3)$.

Let us quote here  
the result for $ b_{m,ny} $ in the case of 
scalar,  mixed, or  vector BMN operators
in \eqref{three}  \cite{GKT}: 
\beqa
\label{bsc}
\left[ b_{m,ny}  \right]_{\rm scalar}&=& m^2 - {mn\over y} \ , 
\\
\label{bmix}
\left[ b_{m,ny}  \right]_{\rm scalar-vector}&=& 
{1 \over 2} \left( m^2 - {n^2 \over y^2} \right)
\ , 
\\
\label{bvec}
\left[ b_{m,ny}  \right]_{\rm vector}&=& -{n^2\over y^2}  + 
{mn\over y} 
\ . 
\eeqa
We conclude this section with one important comment, 
which anticipates our results for the fermions 
to be derived in the next section:  
we will show that, for fermion BMN operators, 
the coefficients $ b_{m,ny}$ for the various representations 
precisely take one of 
the three expressions \eqref{bsc}-\eqref{bvec}.

\subsection{The three-point function of fermion BMN operators}
\label{sec-fer}
In the previous section we explained how to obtain 
 the matrix elements of the SYM 
dilatation operator
in any arbitrary basis of SYM operators, and 
specifically in the isomorphic to string basis. 
Here we present the field theory computation  
of the coefficients $ b_{m,ny}$ appearing in 
\eqref{42}, from which the coefficient of the 
conformal three-point function 
of two-impurity fermion BMN operators can also be derived.
The matrix elements \eqref{Gammap}
of the SYM dilatation operators 
in the natural string basis
will then be obtained
using the expressions for $ b_{m,ny}$ and 
\eqref{matrixS}-\eqref{matrix-t} and \eqref{a-all}. 
The reader not interested in the details of this calculations 
may skip a few pages, and proceed directly to  section 6.

Let us consider the three-point function
of the operators in \eqref{34}, i.e.
\beq
\label{mam}
\langle\cO_{3\a, 4\b;  n}^{y\cdot J} (x_1)
\, 
\cO_{\rm vac}^{(1-y)\cdot J} (x_2)
\, \bar{\cO}_{3\dot\a, 4\dot\b;  m}^{J} (x_3)
\rangle
\ . 
\eeq
We start off by evaluating the Feynman diagrams 
which  originate from the
pure BMN parts of both the barred and unbarred operators.
These diagrams are represented in Figure 1, 
where we draw only the diagrams where the impurities 
$ \lambda_{4\a}$ and $\bar{\lambda}_{4 \dot\b}$ 
participate in the interactions, and the other impurity 
propagates freely. In diagram 1{\it a} (type I),  
the interacting impurity goes across,  
while in 1{\it b}  the interacting impurity 
goes straight (type II). 
The latter diagram has a
minus sign relative to the  former from the 
Yukawa vertex (see \eqref{lag}).
\begin{figure}[ht]
\label{figure1}
\psfrag{l3a}{\LARGE${\bar\l_{3\dot\a}}$}
\psfrag{l4a}{\LARGE${\bar\l_{4\dot\b}}$}
\psfrag{ll4a}{\LARGE${\l_{4\a}}$}
\psfrag{zb}{\LARGE${\overline{Z}}$}
\psfrag{x1}{\LARGE${x_1}$}
\psfrag{x2}{\LARGE${x_2}$}
\psfrag{1a}{\LARGE${1a}$}
\psfrag{1b}{\LARGE${1b}$}
\psfrag{1c}{\LARGE${1c}$}
\psfrag{1d}{\LARGE${1d}$}
\psfrag{1e}{\LARGE${1e}$}
\psfrag{1f}{\LARGE${1f}$}
\begin{center}
\scalebox{0.55}{\includegraphics{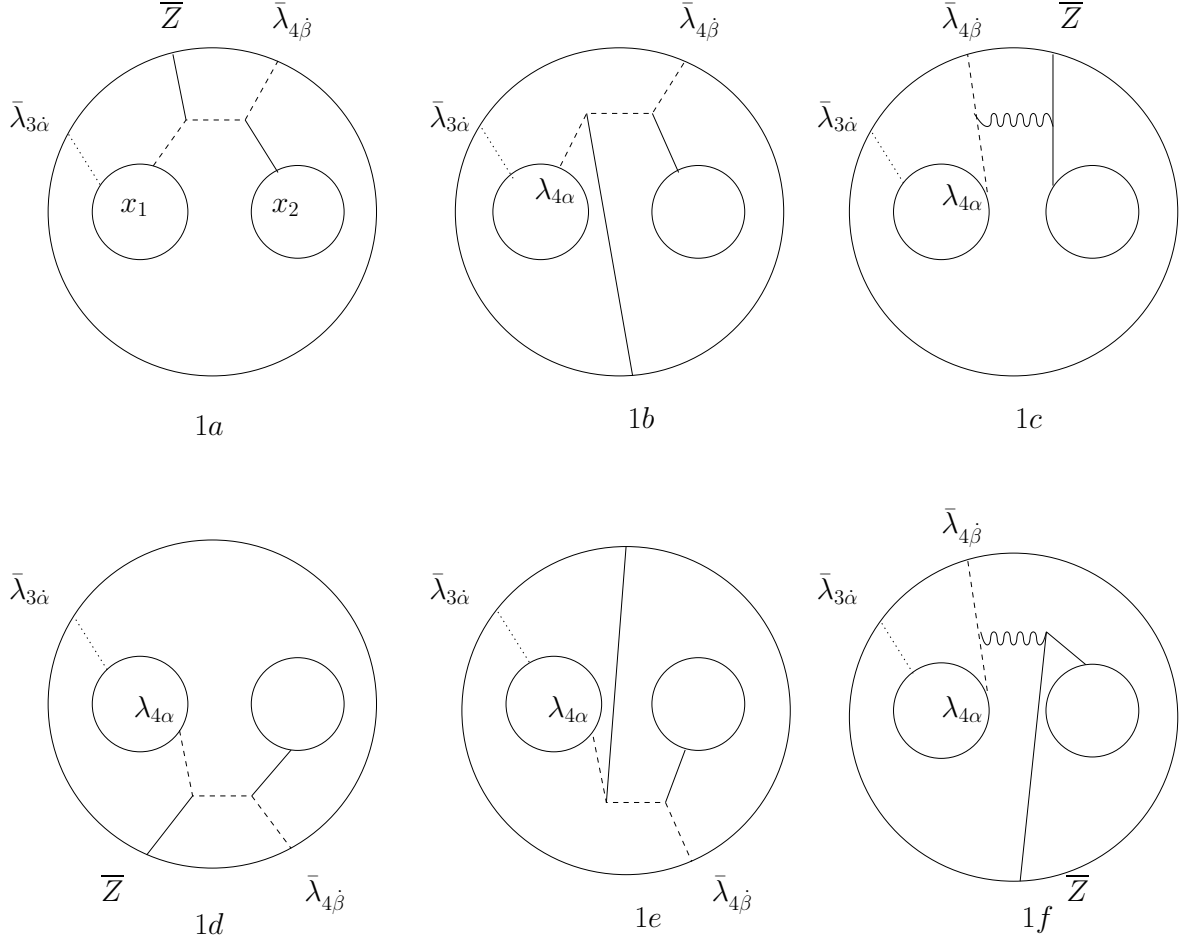}}
\end{center}
\caption{\it Feynman diagrams from the pure BMN parts: of type I (in 1a and 1d) and type II (1b and 1e). 
Diagrams 1d, 1e, are the mirrors of 1a, 1b. 
Diagrams 1d and 1e have phase factors 
which are the complex conjugate of those of 1a and 1b.
The gluon interaction diagrams in  1c and 1f 
have the same BMN factor and 
cancel each other.
}
\end{figure}
The result for the type I diagram
is, concentrating on the interacting part:
\bea
\label{J^I}
J^I_{\a \dot\b}=
(\partial^1_{\nu} \s^{\nu}_{\a \dot\a}) (-\sqrt{2} i
\epsilon^{\dot\psi \dot\a})  
\s^{\rho}_{\chi \dot\psi}(-\sqrt{2} i
\epsilon^{\b \chi})(-\partial^3_{\mu} 
\s^{\mu}_{\b \dot\b}) H_{\r 1423}
\ , 
\eea
where
\bea
H_{\r 1423}&=& 
\int\! d^4x \, d^4y \
\D(x_1-x) \D(x_4-x) \bigl[-\partial^x_{\r}\D(x-y)\bigr]
\D(x_3-y) \D(x_4-y) 
\nonumber\\
&=& -(\partial^1_{\r}+\partial^4_{\r})H_{1432}
\ . 
\eea
Notice that in order to be able to perform the inversion 
in the conjugated operator, we momentarily split 
the insertion points of the fermion  and the $Z$ 
impurity, respectively $x_3$ and $x_4$.
The last equality is obtained after integrating 
by parts with respect to $x$
and then converting the $x$ derivative acting on 
$\D(x_1-x)$ and $\D(x_4-x)$
to derivatives with respect to $x_1$ and $x_4$,  respectively.
The partial derivatives in  \eqref{J^I} come from the fermion
propagator 
$S_{\a \dot\a}(x)=-\partial_{\a \dot\a}\D(x)$, 
where $\partial_{\a \dot\a} :=\partial_\m \s^{\m}_{\a \dot\a}$. 

Using now \eqref{sss} 
and ignoring the $\e$ term which eventually does not 
contribute to  the $\log x_{12}^2 $ term, one obtains:
\bea
J^I_{\a \dot\b}\, =\, 
2\big[ \partial^{3}_{\a \dot\b}J_A
+ \partial^{1}_{\a \dot\b}J_B+
 (\partial^{1}+\partial^{4})_{\a \dot\b}J_C
\bigr]
\ , 
\eea
where 
\bea
\label{JA}
J_A&=&\partial^1 \cdot(\partial^1+\partial^4)\, H_{1423}
\ , 
\\
\label{JB}
J_B&=&\partial^3 \cdot(\partial^1+\partial^4)\, H_{1423}
\ , 
\\
\label{JC}
J_C&=&-\partial^3 \cdot\partial^1 \, H_{1423}
\ . 
\eea
The explicit expressions for $J_A, J_B$ and $J_C$ are 
worked out in 
\eqref{6A}-\eqref{6C}. 
After some algebra,  one  realises that the only non-zero
contribution to $J^I_{\a \dot\b}$ comes from the term involving $J_A$.
Keeping track of the relevant to us terms which contain  
$\log x_{12}^2$,  we get:
\bea
\label{prev}
J^I_{\a \dot\b}\ = \ \frac{1}{2^3 \pi^2} \log x_{12}^2
\, \D(x_4-x_1)\,
\bigl[ \partial^4_{\a \dot\b}\D(x_4-x_1) \bigr]
\ . 
\eea
We note that  \eqref{prev} is precisely of  the form of 
(a first-order correction to) two freely propagating
fields, one $Z$ boson and one fermion, as it is expected.
In order to make the comparison with the string amplitude, 
we should now apply the inversion 
on the 
conjugated operator,  
that is on the scalar $Z$ field and the interacting fermion. 
For the  scalar  field this is rather trivial:  
according to \eqref{ccc}, one has to multiply $J^I_{\a \dot\b}$ 
by $x_4^2$. 
For the fermion, \eqref{Weyl} instructs us to 
multiply by $\eta x_4^2\bar\s_{\m}^{\dot\b \b}x_4^{\m}$.
Taking into account the identity 
$(\s_{\mu})_{\a \dot\b}\bar\s_{\nu}^{\dot\b \b}
x^{\mu}x^{\nu}=\d_{\a}^{\b}\, x^2$ one  obtains:
\beqa
\label{lw}
 - 4 \Big(\frac{g^2}{2}\Big)^3 P_I
\frac{\log x_{12}^2}{2^8\p^6}\, \eta^2\d_{\a}^{\b}
\ .
\eeqa
The overall factor $({g^2}/{2})^3$ comes from
the insertion of two vertices, which give $({2}/ {g^2})^2$,
and five propagators, which give $({g^2}/ {2})^5$. 
$P_I$ is the phase factor associated with the diagrams 
of type I, and is explicitly calculated in \eqref{fFf}.
In order to obtain the final result for the type I diagrams, 
we have still to multiply \eqref{lw} by a factor of $2$ from
the free contraction of the non-interacting impurity, 
and by a factor of $1/4$ from the normalisation
of the two fermion BMN operators. Doing so, and setting
$\eta^2=1$ we get:
\beqa
{\rm type \ I}-{\rm fermions:} &&
 -2 \Big(\frac{g^2}{2}\Big)^3 P_I
\frac{\log x_{12}^2 }{2^8\p^6}\d_{\a}^{\b}
\ . 
\eea
Notice that this result is precisely the same as the result 
for the case of two different scalar impurities. 
Similarly, one gets, for the type II diagrams:
\beqa
{\rm type \ II}-{\rm fermions:} &&
2 \Big(\frac{g^2}{2}\Big)^3 P_{II}
\frac{\log x_{12}^2}{2^8\p^6}\d_{\a}^{\b}
\ . 
\eeqa
To the  type II diagrams is associated the 
phase factor $P_{II}$ of \eqref{fFf}. Moreover, 
in order to get the final expression for 
the coefficient $b_{m,ny}$, 
one should also include the diagrams where 
the other impurity, $\l_3$, participates in the interaction.

If this was the whole story, we would conclude 
that three-point functions of fermions take the same form 
as the three-point functions for scalars. 
But the correct expressions for BMN operators often  contain 
compensating  terms, and so is the case for 
the operator in \eqref{34}. 
Importantly, these compensating terms do contribute to 
the  three-point functions and must be taken into account.
In our case, the compensating terms affect the
fermion  operators that 
have a projection in the ${\bf (1,3^+)}$ representation 
(see the discussion after \eqref{decT}), and 
the corresponding contribution
is important and we will now compute it.

In Figure 2 we draw the Feynman diagrams
obtained by taking  the compensating term 
on the right hand side of \eqref{34}
for the operator sitting at $x_1$  . 
To the diagrams 2{\it a} and 2{\it c} 
a BMN phase factor equal to 1 is associated.
\begin{figure}[ht]
\label{figure2}
\psfrag{l4b}{\LARGE${\bar\l_{3\dot\a}}$}
\psfrag{l3a}{\LARGE${\bar\l_{4\dot\b}}$}
\psfrag{2a}{\LARGE${2a}$}
\psfrag{2b}{\LARGE${2b}$}
\psfrag{2c}{\LARGE${2c}$}
\psfrag{2d}{\LARGE${2d}$}
\begin{center}
\scalebox{0.45}{\includegraphics{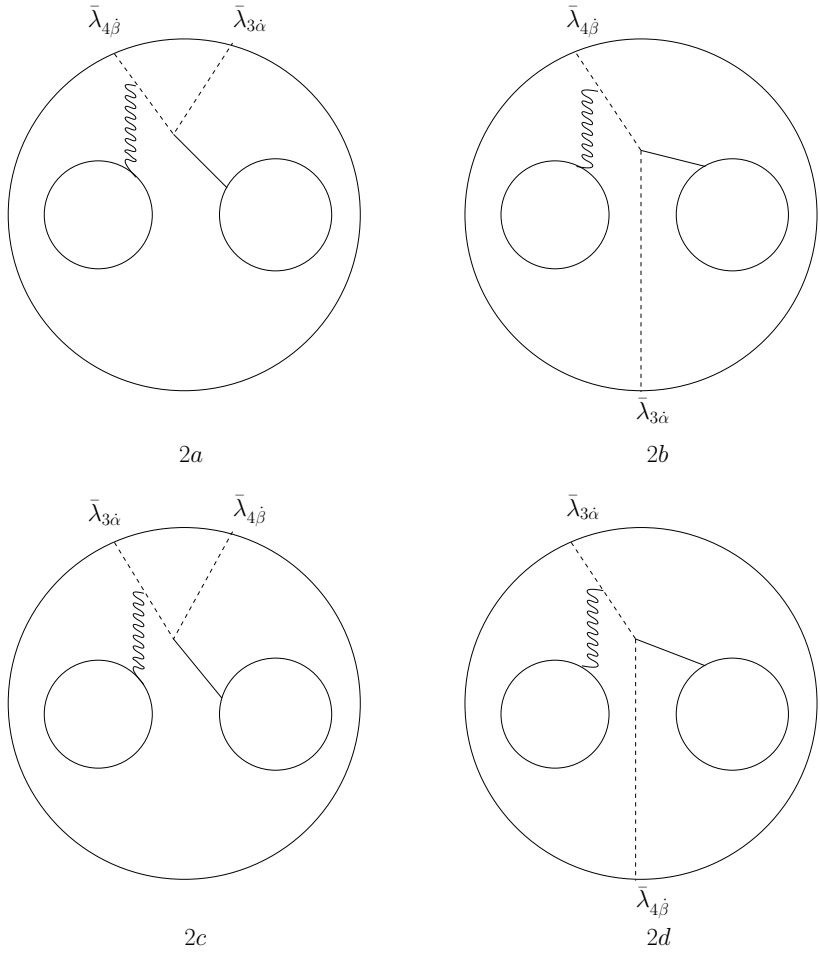}}
\end{center}
\caption{\it Gluon emission diagrams originating from the compensating term in the internal operator.
The gluon is absorbed by the fermion field. 
There are also mirror diagrams, not drawn in this figure.}
\end{figure}
{}From the first diagram in Figure 2 we get:%
\footnote{As in \cite{GKT}, 
the diagrams where the compensating term is taken in 
the external operator, or both in 
the external and internal operator, 
do not contribute.}
\bea
\label{Ifirst}
I_{\hat\a\hat\gamma,\dot\a\dot\b}\ =\ 
-2 \cdot(-1)^3 \partial_{\m}^4 \s^{\m}_{\b \dot\b}
\partial_{\n}^3 \s^{\n}_{\a \dot\a}(-i \bar\s_{\rho}^{\dot\gamma \b})
\s^{\tau}_{\gamma\dot\gamma} (-\sqrt{2}i \epsilon^{\gamma\a})
\partial_{\d}^1(\s_{\d\rho })_{\hat\gamma}^{\hat\b}\e_{\hat\a\hat\b}
(-H_{\tau 1432})
\ . 
\eea
The factor of $(-1)^3$ comes from three propagators, 
while the
factor of $2$ arises from the two terms in the 
field strength  $F_{\rho\d}$. 
The $\s_{\rho\d}$ is related to the compensating term 
of the operator at $x_1$ while the
$\bar\s_{\rho}$ matrix comes from the 
gluon-fermion interaction vertex.
Finally, the minus sign in front of \eqref{Ifirst}
comes from Wick contracting fermions.

We can now elaborate \eqref{Ifirst} using the 
completeness relation  \eqref{completeness},  
to obtain:
\bea
I_{\hat\a\hat\gamma,\dot\a\dot\b}=
2\sqrt{2}\big[
(\s^{\d}\bar\s^{\tau}\s^{\nu})
_{\hat\gamma\dot\a}\s^{\mu}_{\hat\a\dot\b}\, + \, 
(\s^{\d}\bar\s^{\tau}\s^{\nu})_{\hat\a\dot\a}
\s^{\mu}_{\hat\gamma\dot\b}\big]
\, 
\partial_{\m}^4\partial_{\n}^3
\partial_{\d}^1\, 
(-H_{\tau,1432})
\ .
\eea
\begin{figure}[ht]
\label{figure3}
\psfrag{l3a}{\LARGE${\bar\l_{3\dot\a}}$}
\psfrag{l4b}{\LARGE${\bar\l_{4\dot\b}}$}
\psfrag{3a}{\LARGE${3a}$}
\psfrag{3b}{\LARGE${3b}$}
\psfrag{3c}{\LARGE${3c}$}
\psfrag{3d}{\LARGE${3d}$}
\begin{center}
\scalebox{0.45}{\includegraphics{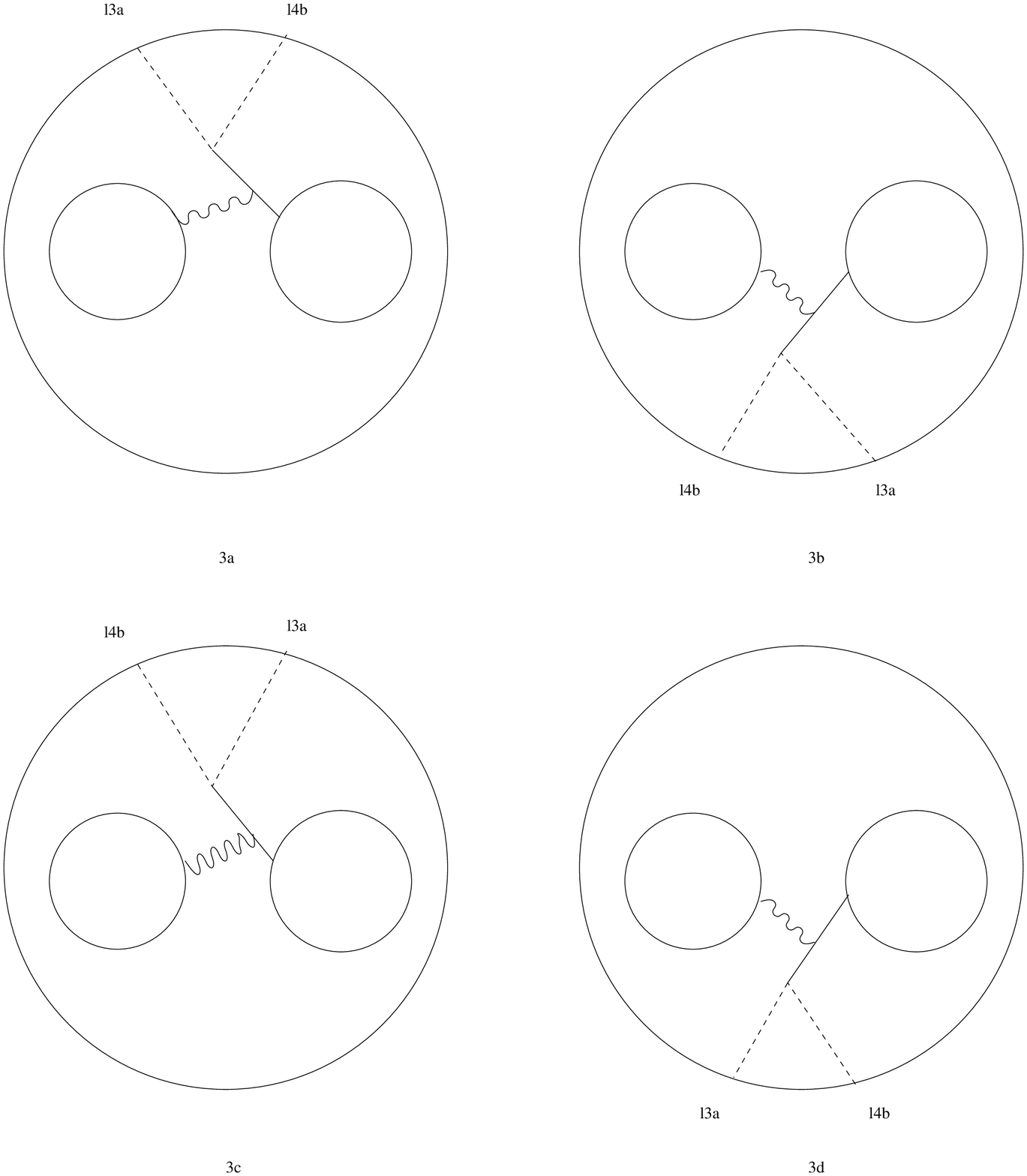}}
\end{center}
\caption{\it Gluon emission diagrams originating
from the compensating term in the internal operator.
The gluon is absorbed by the $Z$ field.}
\end{figure}

\noindent
Using \eqref{sss}, and discarding the irrelevant
$\epsilon$-terms as before, we get:
\bea
I_{\hat\a\hat\gamma,\dot\a\dot\b}
=2\sqrt{2} {\partial^4_{\hat\a\dot\b}}
\big(  {\partial^3_{\hat\gamma\dot\a}} J_A+
 {\partial^1_{\hat\gamma\dot\a}} J_B+
 {(\partial^1+\partial^4)_{\hat\gamma\dot\a}}J_C\big) +
\hat\a \longleftrightarrow\hat\gamma
\, .
\eea
Similarly to the diagrams in Figure 1,  
the  $\log x_{12}^2$ contributions from 
$J_B$ and $J_C$ cancel out, 
and we are left with:
\bea
I_{\hat\a\hat\gamma,\dot\a\dot\b}\ =\ 
8\sqrt{2}\frac{1}{2^8 \pi^6} \log x_{12}^2
\, 
(\s^{\mu}_{\hat\a\dot\b}\s^{\nu}_{\hat\gamma\dot\a}+
\s^{\mu}_{\hat\gamma\dot\b}\s^{\nu}_{\hat\a\dot\a})
\frac{x_3^{\mu}x_3^{\nu}}
{(x_3^2)^4}
\ . 
\eea
Now we perform the inversion for the fermions, thus  arriving at:
\bea\label{I}
I_{\hat\a\hat\gamma}^{\a\b}\ = \ 
8\sqrt{2}\frac{1}{2^8 \pi^6} \log x_{12}^2 \, 
(\d^{\a}_{\hat\a}\d^{\b}_{\hat\gamma}\ + \ 
\d^{\a}_{\hat\gamma}\d^{\b}_{\hat\a})
\ . 
\eea
The calculation for the diagram in Figure 2{\it c} 
(where the other fermion absorbs the gluon) proceeds 
in a similar fashion to that of Figure 2{\it a}, 
giving the same result 
as in \eqref{I} but with $\a$ and  $\b$ swapped.
Since the expression for $I_{\hat\a\hat\gamma}^{\a\b}$ 
is symmetric under this exchange,  
the results is the same as 
for the diagram in Figure 2{\it a}. 
We should also not forget to multiply our result 
by $1/4$ due to
the normalisation of the operators ($1/2$ for each operator), 
and by a factor of 
$-\sqrt{2}/4$ coming from 
the compensating term
on the right hand side of \eqref{34}.
Taking also  account the powers of $g^2/2$ associated 
with vertices and propagators,  we finally get 
the result for the sum of diagrams 2{\it a} and 2{\it c}:
\bea
\label{abbb}
-2\left(
\frac{g^2}{2}\right)^3
\frac{1}{2^8 \pi^6} \log x_{12}^2  \, 
(\d^{\a}_{\hat\a}\d^{\b}_{\hat\gamma}\ +\ 
\d^{\a}_{\hat\gamma}\d^{\b}_{\hat\a})
\ . 
\eea
The diagrams in Figure 2{\it b} and 2{\it d}
can be computed in a similar way. Notice that they 
have a relative minus sign compared to 
2{\it a} and 2{\it c}, and a different BMN factor
$\bar{q}^{J_1 + 1} $. 

For $\a\neq \b$, the result of 
\eqref{abbb} is exactly the same as the result 
obtained from the compensating diagrams
for  the case of BMN operators 
with {\it mixed} impurities (one vector and one scalar impurity). 
We addressed this case in section 4 of  \cite{GKT}.
Furthermore, for $\a= \b$ we get the same result as that
of the compensating diagrams of BMN operators 
with {\it vector} impurities, i.e.~twice that of the mixed case 
(see  \cite{CKTvec,GKT}).
{}From \eqref{bsc}-\eqref{bvec}, one can easily work out 
the contributions of the compensating terms alone
for the mixed and vector case, 
\beqa
\left[ b_{m,ny}  \right]_{\rm mixed}^{\rm c.t.}&=&
\left[ b_{m,ny}  \right]_{\rm mixed} \ - \ 
\left[ b_{m,ny}  \right]_{\rm scalar}\ = \ 
-{1\over 2}\Bigl( m-\frac{n}{y}\Bigr)^2 
\ , 
\\ 
\left[ b_{m,ny}  \right]_{\rm vector}^{\rm c.t.}&=&
\left[ b_{m,ny}  \right]_{\rm vector} \ - \ 
\left[ b_{m,ny}  \right]_{\rm scalar} \ = \ 
- \Bigl(m-\frac{n}{y}\Bigr)^2
\\ \nonumber   
&=& 2 \, \left[ b_{m,ny}  \right]_{\rm mixed}^{\rm c.t.}
\ .
\eeqa

Finally, let us note that at the order we are working, 
the diagrams of Figure 3 are also present. However,
the diagram in 3{\it a} cancels against that in 
3{\it b} because of the relative minus sign associated
with the vertex where the gluon is absorbed; 
and, similarly, 3{\it c} cancels against
3{\it d}. Therefore  
the net contribution of the diagrams in Figure 3
is zero.

We conclude by summarising the results of this section. \\
{\bf a.} The contribution of the pure BMN part of the operators 
involved in the three-point function 
\eqref{mam} is precisely the same obtained for 
{\it scalar} BMN operators.
Hence the corresponding coefficient 
$b_{m,ny}$ is given by \eqref{bsc}.\\
{\bf b.} The previous remark also implies that, 
when no compensating term is present in the expression for the 
BMN operator considered, 
the result \eqref{bsc} gives the full answer. \\
{\bf c.} When a compensating term appears, 
it contributes precisely as the compensating term 
of the mixed (scalar-vector) case when $\a\neq \b$, 
of the vector case for $\a = \b$. 
The corresponding expressions for $b_{m,ny}$
are then \eqref{bmix} and \eqref{bvec}, respectively.

\section{Testing the BMN correspondence in the fermion sector}
We can now apply the results derived in the previous section  
to test the pp-wave/SYM correspondence in the fermionic sector. 
In particular, we will  
reproduce  in the gauge theory   
the three-string amplitudes for the
following flavour-conserving processes:
\bea
\label{process1}
(\l_{31}\ldots\l_{32})_m\, 
\longrightarrow\,  (\l_{31}\ldots\l_{32})_n
\, +\, {\rm vac.} \ , 
\\
\label{process2}
(\l_{31}\ldots\l_{31})_m\, 
\longrightarrow\,  (\l_{31}\ldots\l_{31})_n
\, +\, {\rm vac.} \ , 
\\
\label{process3}
(\l_{31}\ldots\l_{42})_m\, 
\longrightarrow\,  (\l_{31}\ldots\l_{42})_n
\, +\, {\rm vac.} \ , 
\\
\label{process4}
\!\!(\l_{31}\ldots\l_{41})_m\, 
\longrightarrow \,  (\l_{31}\ldots\l_{41})_n
\, +\, {\rm vac.} \ .
\eea
A few comments are in order. 
\begin{enumerate}
\item
First, 
it does not take long to realise that 
these cases  actually cover 
all the irreducible representations 
of the two-impurity fermion BMN operators,  
where the two impurities are Weyl spinor of the same chirality
(see \eqref{decomp1} and \eqref{decomp2}).
\item
The case where the two impurities have opposite chirality,  
$\lambda_{r\a}$ and $\bar\lambda_{\dot{r} \dot\b}$
(see \eqref{decomp3}),  
can of course be treated with a similar technique. 
Notice that, in that case, a compensating term 
containing ($D_{\m} \phi^i)\sigma_{\a \dot\b}^{\m}$ 
will always be present in the precise 
expression of the BMN operator containing 
the $\lambda_{r\a}$, $\bar\lambda_{\dot{r} \dot\b}$ impurities, 
in agreement with the 
fact that the right hand side of 
\eqref{decomp3} 
contains only one irreducible representation.
\item
Finally, notice that the operators taking part 
in the first two processes 
\eqref{process1}   and \eqref{process2}    
do not have compensating terms, 
i.e.~they do not have a projection onto 
the ${\bf (1,3^+})$ representation
of $SO(4)\times SO(4)$, in contradistinction with 
the operators taking part in the remaining last two, 
\eqref{process3}   and \eqref{process4}.    
\end{enumerate}
 
Let us write down the string amplitudes 
corresponding to the processes of 
\eqref{process1}-\eqref{process4},  
taking into account that in 
the three-string vertex of 
\cite{SV1,SV2,PS,Panknew}
all the external states are written as ket states:
\bea
\label{ampl.}
\m^{-1}
\langle {\rm v}|\, \b_{-m(3)}^{12}\, \b_{m(3)}^{11}
\, \b_{11,n(1)}\, \b_{12,-n(1)}\, |H_3\rangle &\equiv& 
A_1 \ , 
\nonumber\\
\m^{-1}
\langle {\rm v}|\, \b_{-m(3)}^{11}\, \b_{m(3)}^{11}
\, \b_{11,n(1)}\, \b_{11,-n(1)}\, |H_3\rangle &\equiv & A_2
\ , 
\nonumber\\
\m^{-1}
\langle {\rm v}|\, \b_{-m(3)}^{22}\, \b_{m(3)}^{11}
\, \b_{11,n(1)}\, \b_{22,-n(1)}\, |H_3\rangle& \equiv & A_3
\ , 
\nonumber\\
\m^{-1}
\langle {\rm v}|\, \b_{-m(3)}^{21}\, \b_{m(3)}^{11}
\, \b_{11,n(1)}\, \b_{21,-n(1)}\, |H_3\rangle& \equiv&  A_4
\ . 
\eea
We have already computed the string amplitudes 
in \eqref{ampl.} in  \eqref{res1}-\eqref{res4}, finding 
\beqa
\label{Res1}
A_1 &=& 
- \, \l' \, 
C_{\rm norm}\frac{\b+1}{\pi^2 }\sin^2{\pi m  \b} \ , 
\\
\label{Res2}
A_2 &=& 0 \ , 
\\
\label{Res3}
A_3 & =& 0 \ , 
\\
\label{Res4}
A_4 &=& 
\l' \, 
C_{\rm norm}\frac{\b+1}{\pi^2 }\sin^2{\pi m \b} 
\ . 
\eeqa

We start our analysis from 
the first process \eqref{process1}. 
For this case,   we have found in the previous section that 
the corresponding coefficient $b_{m,ny}$
of the field theory three-point function 
is exactly the same as that 
obtained in the case of BMN operators
with two scalar impurities of different flavours. 
This is due to the absence  of compensating terms 
in the operators participating in the process 
\eqref{process1}, 
so that only   the diagrams of Figure 1 contribute
(with $\l_4$ replaced by the second $\l_3$). 
As explained in  \eqref{a-all}, 
supersymmetry guarantees that the anomalous dimension 
of the two fermion BMN operators 
is the same as that of two scalars;   
therefore, the coefficient $a_{m,ny}$ 
for the fermions is identical to that of 
two different scalars. 
The consequence of this is that the gauge theory 
prediction for the string amplitude
of \eqref{process1}  is exactly the same
as the  prediction obtained in the case of BMN operators with 
two different scalar impurities; 
and it was shown in \cite{GKT}
that there  is precise agreement between 
the field theory and string theory prediction 
in the case of two scalar impurities of different flavours.
The result obtained in string field theory 
for the process \eqref{process1} is given in \eqref{Res1}. 
Indeed, this result \eqref{Res1} precisely coincides with 
the string amplitude for the case of 
two scalar impurities, Eq.~\eqref{sv1}.
This is our first test.

Two identical fermions, 
both in the unbarred and barred  operators, 
take part in the  process in  \eqref{process2}. 
The corresponding gauge theory calculation is therefore 
slightly more complicated, since there are twice as many
contractions  as in the previous case, 
and thus twice as many Feynman diagrams. 
Taking these diagrams into account, 
the $S$ and $T$ matrices take the following form:
\beqa
S &=&
\left(\begin{tabular}{cc}
$\d_{mn}$ & $g_2\, ( C_{m,qz}- C_{m,-qz})$  \cr \cr
$ g_2 \,( C_{py,n}-C_{-py,n})$ &  $\d_{pq}$
\end{tabular}\right) \ + \ \cO (g_2^2) 
\\ \nonumber \cr \cr
& =&   \uno + g_2 s +\cO (g_2^2)  \ ,
\\ \cr \cr 
T &=& \l' \left(\begin{tabular}{cc}
\label{514}
$ m^2\, \d_{mn}$ & $  g_2\,\big[ C_{m,qz} \, ( a + b)_{m,qz}$  \cr
$ $ & $-C_{m,-qz} \, ( a + b)_{m,-qz}\big]$ \cr \cr
$ g_2\, \big[ C_{py,n}\, (a+b)_{py,n}$  & 
$(p^2 / y^2)\, \d_{pq} \d_{yz}$\cr
$- C_{-py,n}\, (a+b)_{-py,n}\big]$
\end{tabular}\right) \ + \ \cO (g_2^2)
\\   \cr \cr 
&=& d\  + \ g_2 t  \ + \ \cO (g_2^2)
\ .
\nonumber
\eeqa
In \eqref{514} the coefficient 
$a_{m,ny}$ is as in \eqref{a-all}, and 
$b_{m,ny}$  is  given in \eqref{bsc}, as explained 
in section 5.
We should note the crucial minus sign between the two terms
appearing in the non-diagonal matrix elements of $T$. This
comes from the anticommuting nature of the fermion impurities.
We can now work out the expression for the 
matrix $\Gamma$
of the SYM dilatation operator in the field theory basis 
which is isomorphic to the natural string basis.
Using \eqref{Gammap}, we get immediately 
\beqa
\Gamma=d+g_2\Bigl[ t-(1/2)\{s,d\} \Bigr]=\l' \left(\begin{tabular}{cc}
$ m^2\, \d_{mn}$ & $ 0$ \cr \cr
$0 $ & $(p^2 / y^2)\, \d_{pq} \d_{yz}$
\end{tabular}\right)
\ . 
\eeqa
Thus, we conclude that the field theory prediction
for the second process \eqref{process2} is $0$. This is 
in agreement with the vanishing of the corresponding 
string  amplitude of \eqref{Res2}.

Next, we consider the third process, \eqref{process3}. 
In this case the compensating diagrams of Figure 2 
should be taken into account.
As we have noticed in the previous section 
(see the discussion after \eqref{abbb}), 
the contribution of the compensating term  is the same
as that arising from compensating terms of 
BMN operators with mixed (one scalar-one vector) 
impurities.
Furthermore, the contribution of the diagrams  
where only the pure BMN parts of the operators 
is taken into account,  
is the same for all the cases (i.e.~scalar,
mixed, vector and fermion impurities). 
Therefore, we conclude that the coefficient $b_{m,ny}$
appearing in the the three-point function, and thus
the whole calculation, are  
identical to the mixed case
studied in \cite{GKT}.%
\footnote{As before, in order to reach this conclusion we
also used the fact that the anomalous dimension 
of all  two-impurity operators, i.e.~is the $a_{m,ny}$ 
coefficient,  is the same for any kind of impurity.}
It was found in \cite{GKT} that 
the matrix elements of the dilatation operator
in the isomorphic to string basis for the case of BMN
operators with mixed impurities is given by 
\beq
\label{macc}
\Gamma_{\rm mixed} \ = \ 
d+g_2\Bigl[t_{\rm mixed} -(1/2)\{s,d\}\Bigr]\ =\ 
\l' \left(\begin{tabular}{cc}
$ m^2\, \d_{mn}$ & $ 0$ \cr \cr
$0 $ & $(p^2 / y^2)\, \d_{pq} \d_{yz}$
\end{tabular}\right)
\ . 
\eeq
Therefore, the previous result 
\eqref{macc} precisely reproduces, in the gauge theory, 
the vanishing three-string
amplitude of \eqref{Res3}.

Finally, we focus on the last process of  \eqref{process4}.
We noticed in section 5  that the diagrams from the 
compensating terms  contribute, in this case,   
exactly as the diagrams from the compensating term 
for two vector operators. 
Following similar arguments as before, 
we conclude that the field theory
prediction for this process is 
the same as that for the vector case. 
The result for the vector case was found in \cite{GKT}
to be equal to the negative of 
the result for  the process 
for the scalars, which in turns 
is equal to   the result for \eqref{process1}.
This is again in perfect agreement 
with the string amplitude obtained  
in \eqref{Res4}.  This is our last test.

We close this section with a comment about how the ${\bb Z}_2$ 
symmetry of the pp-wave background is realised 
in the string amplitudes of \eqref{ampl.}.
It is known that under the  ${\bb Z}_2$ symmetry 
the two indices of a fermion
creation or annihilation operator are exchanged, 
\bea
\label{z2}
{\bb Z}_2: \qquad \b_{\a\b}\longrightarrow \b_{\b\a}
\ . 
\eea
However, whereas the string vertex 
$|H_3\rangle$ is invariant under ${\bb Z}_2$,
the true vacuum $|{\rm v} \rangle$ 
corresponds to a combination of the trace of the metric
and the five-form field on one 
of the $\bb{R}_4$'s \cite{MT},   
and thus one has to assign
negative ${\bb Z}_2$ parity to it 
\cite{Panknew,KKPR}. The correctness of this assignment was also 
verified from the field theory perspective in \cite{GKT}
(see also the discussion in the Introduction). 
If we apply \eqref{z2} to the string amplitudes
in \eqref{ampl.} we obtain:
\bea
\hat {\bb Z}_2 A_1&\equiv&  -A_4= A_1
\ , 
\nonumber\\
\hat {\bb Z}_2 A_2 & \equiv&  -A_2= A_2
\ , 
\nonumber\\
\hat {\bb Z}_2 A_3 & \equiv & -A_3= A_3
\ , 
\nonumber\\
\hat {\bb Z}_2 A_4 & \equiv & -A_1= A_4
\ . 
\eea
Therefore, we conclude that  the $\bb{Z}_2$ 
symmetry leaves the value of the string amplitudes of 
\eqref{process1}-\eqref{process4} invariant.



\newpage
\section*{Acknowledgements}
It is a pleasure to thank Carlo Angelantonj, Andreas Brandhuber,
Sanjaye Ramgoolam, Rodolfo Russo, 
Bill Spence and in particular  Valya Khoze
for stimulating conversations.
GG acknowledges a grant from the State Scholarship Foundation of
Greece (I.K.Y.), and the hospitality of Queen Mary College,
London, where part of this work was carried out. 
The work of GT was supported by PPARC.


\newpage

\startappendix
\Appendix{Notation and conventions in gauge theory}
\label{appendixnc}
We write the Euclidean $\N =4$ Lagrangian as
\bea
{\cal{L}}={\cal{L}_B}+{\cal{L}_F}
\ , 
\eea
where
\bea
{\cal{L}_B} = {2\over g^2}  \Tr \left( {1 \over 4}
F_{\mu \nu}F_{\mu \nu} +
(D^\mu \bar Z^i ) (D_\mu Z_i  )-[Z_i,Z_j][\bar Z^i,\bar Z^j]
+{1\over 2}[Z_i,\bar Z^i][Z_j,\bar Z^j]\right)
\ , 
\eea
and
\beqa
\nonumber
\label{lag}
{\cal{L}_F}={2\over g^2} \Tr \left(  \l_{A} \s^{\mu} D_{\mu} \bar\l^A
-\sqrt{2} i ([\l_4,\l_i]\bar Z^i+[\bar\l^4,\bar \l^i]Z_i)
+{i \over \sqrt{2}}(\epsilon^{ijk}[\l_i,\l_j]Z_k
+\epsilon_{ijk}[\bar\l^i,\bar\l^j]\bar Z^k)\right)
\ . 
\\
\eeqa
In the above equation $A=1,\ldots , 4$ and $i,j,k=1, \ldots , 3$.
$Z_i$ are the the three complex scalars  defined by
\beq
\label{compbas}
Z_1  = {\phi_1 + i \phi_2 \over \sqrt{2}} \ , \quad
Z_2  = {\phi_3 + i \phi_4 \over \sqrt{2}} \ , \quad
Z_3  = {\phi_5 + i \phi_6 \over \sqrt{2}} \ ,
\eeq
where  $\phi_i$, $i=1, \ldots , 6$  are the real scalar fields
transforming  under the $SO(6)$ R-symmetry group.
We will also set $Z_3:=Z$.

We define the  covariant derivative is
 $D_\mu \phi_i = \partial_{\mu}\phi_i  - i [A_\mu, \phi_i ]$,
where
$A_\mu = A_\mu^{a} T^a$, and $F_{\mu \nu} = \partial_{\mu}A_{\nu} -
\partial_{\nu}A_{\mu} - i [ A_{\mu} , A_{\nu}]$.


Our $SU(N)$ generators are normalised as
\beq
\Tr \left( T^a T^b \right) = \delta^{ab} \ ,
\eeq
so that, for example,
\beqa
\left< Z^{i}_{j}(x) \bar{Z}^{l}_{m}(0) \right> =
{g^2\over 2} \, \delta^{i}_{m} \delta^{l}_{j}\, \Delta (x)
\ \ , \ \ \Delta (x)= {1\over 4\pi^2 x^2 } \  .
\eeqa
Our Euclidean sigma matrices 
satisfy
\bea
\s_{\mu}\bar\s_{\nu}+\s_{\nu}\bar\s_{\mu}=2\d_{\mu\nu}
\ ,
\qquad
\bar{\s}_\m \s_\n + \bar{\s}_\n \s_\m = 2 \d_{\m \n}
\ .
\eea
The completeness relation reads:
\bea
\label{completeness}
\s^{\mu}_{\a\dot\b}\, 
\bar\s_{\mu}^{\dot\gamma\d}\ =\ 2\, \d_{\a}^{\d}\, 
\d_{\dot\b}^{\dot\gamma}
\ . 
\eea
Another useful identity is:
\bea\label{sss}
\s_{\nu}\, \bar\s_{\rho}\, \s_{\mu}
\ =\ 
\d_{\nu\rho}\s_{\mu}\, +\, \d_{\mu\rho}\s_{\nu}\, -\, 
\d_{\mu\nu}\s_{\rho}\, +\, \e_{\nu\rho\mu\tau}\s_{\tau}
\eea
We also define $\s_{\mu \nu}$ and $\bar{\sigma}_{\m \n}$ by:
\beqa
\s_{\mu \nu}&=& \frac{1}{2}(\s_{\mu}\bar \s_{\nu}-
\s_{\nu}\bar \s_{\mu})
= i \eta^{a}_{\m \n} \s^a \ ,
\\
\bar{\s}_{\mu \nu}&=&\frac{1}{2}(\bar{\s}_{\mu}\s_{\nu}-
\bar{\s}_{\nu}\s_{\mu})
= i \bar{\eta}^{a}_{\m \n} \s^a \ ,
\eeqa
where $\eta^a_{\m \n}$ ($\bar{\eta}^a_{\m \n}$) 
are the (anti-)self-dual 't Hooft symbols
\cite{'tHooft:fv}.

Finally, we will use the definitions 
$J:= J_1 + J_2$ and $J_1 = y\cdot J$,
where $y\in (0, 1)$.

\Appendix{The three-string vertex}
We begin by specifying the notation and conventions
used in pp-wave string field theory.
The combination $\a'p^+$ for the r-th string is denoted $\a_\rr$
and $\sum_{\rr=1}^3 \a_\rr =0$. As is standard in
the literature, we will choose a frame in which $\a_3=-1$,
\be \label{frame}
\a_\rr = \a'p^+_{(\rr)} \, : \qquad \a_3=-1, \qquad \a_1=y, \qquad \a_2=1-y.
\ee
In terms of the $U(1)$ R-charges
of the BMN operators in the gauge theory
three-point function, $\langle \cO_1^{J_1}\,  \cO_2^{J_2} \, \bar\cO_3^{J}
\rangle$
we have
\be
y=\frac{J_1}{J}\ ,  \qquad 1-y=\frac{J_2}{J}, \qquad y \in (0, 1) \ ,
\ee
and $J=J_1+J_2$.

The effective SYM coupling constant \eqref{lampr} in the frame \eqref{frame}
takes the simple form
\be \label{lamprsim}
 \l' =  \frac{1}{(\mu p^+ \a')^2}\ \equiv \frac{1}{(\mu \a_3)^2}\
= \frac{1}{\mu^2}.
\ee
Here $\mu$ is the mass parameter 
which appears in the pp-wave metric, in
the chosen frame it is dimensionless%
\footnote{It is $p^{+}\mu$ which is
invariant under longitudinal boosts 
and is frame-independent.} 
and the expansion in powers 
of $1/\mu^2$ is equivalent to the perturbative
expansion in $\l'$. Finally,  the frequencies are defined via
\be
\omega_{\rr m}= \sqrt{m^2+(\mu\a_\rr)^2} \ .
\ee

The three-string vertex $\ket{H_3}$ can be represented
as a ket-state in the tensor product of
three single-string Fock spaces.
It has the form \cite{SV1,SV2}
\be
\label{139}
{1\over \m}
\ket{H_3} = {\sf P}  \ket{V_F} \ket{V_B}
\, \d \Big(\sum_{\rr=1}^{3} \a_\rr \Big)
\ ,
\ee
where the kets $\ket{V_B}$ and  $\ket{V_F}$ are constructed
by requiring they satisfy the bosonic and
fermionic kinematical symmetries, and $\a_\rr$ are defined
in \eqref{frame}.
$\ket{V_B}$ is given by
\be
\label{Vb}
\ket{V_B} = \exp\left(
\frac{1}{2} \sum_{\rr,\rs=1}^3 \sum_{m,n=-\infty}^\infty
\sum_{I=1}^8 a^{(\rr)\, I\dagger}_{m}
\overline{N}_{mn}^{(\rr\rs)}  a^{(\rs)\,
I\dagger}_{n}\right) \ket{\rv}_1  \ket{\rv}_2  \ket{\rv}_3   \ ,
\ee
where the $\overline{N}_{mn}^{(\rr\rs)}$ are the Neumann matrices in the
number operator basis.
The fermionic ket state $\ket{V_F}$,  
which is relevant for this paper, is given in the 
$SO(4) \times SO(4)$ formalism  by
\cite{RR,Panknew}
\be
\label{Vf}
\ket{V_F} = \exp\left(
\sum_{\rr,\rs=1}^3 \sum_{m,n \geq 0 }
(
b^{\alpha \beta\, \dagger}_{- m (\rr) } \,
b^{\dagger}_{n (\rs) \, \alpha \beta } \ + \
b^{\dot{\alpha} \dot{\beta}\, \dagger}_{m (\rr)}\,
b^{\dagger}_{-n (\rs) \, \dot{\alpha} \dot{\beta} }
)
\overline{Q}_{mn}^{(\rr\rs)}  \right)
\ket{\rv}_1  \ket{\rv}_2  \ket{\rv}_3   \ ,
\ee
where $\overline{Q}_{mn}^{(\rr\rs)}$ are the fermionic
Neumann matrices.
The complete perturbative expansion of the Neumann matrices
in the pp-wave background
in the vicinity of  $\mu=\infty$,
was constructed in \cite{HSSV}%
\footnote{See also \cite{Lucietti:2004wy}, and Appendix of
\cite{CK} for some useful properties of the 
Neumann matrices.}.
The vacuum state $\ket{\rv} \equiv
\ket{\rv}_1 \ket{\rv}_2  \ket{\rv}_3$
is defined as the state which is annihilated by all
$a$'s and $b$'s,
\beq
a_{n (\rr)} \,  \ket{\rv}_{\rr} \, = \, 0 \ , \ \ \ \
b_{n (\rr)} \,  \ket{\rv}_{\rr} \, = \, 0 \ , \ \ \ \
\forall n \ .
\eeq
The prefactor ${\sf P} $ is determined by
imposing the dynamical symmetries
of the pp-wave superalgebra,
and  was derived in \cite{Panknew}.
Its expressions  reads:
\beqa
\label{pref}
{\sf P} &=& \biggl[
\left({\cK}^i{\widetilde{\cK}}^j\, + \,
\frac{\mu\beta(\beta+1)}{2}
\alpha_3^3\, \delta^{ij}\right)V_{ij}\, - \,
\left({\cK}^a{\widetilde{\cK}}^b\, + \,
\frac{\mu\beta(\beta+1)}{2}\alpha_3^3\, \delta^{ab}\right)V_{ab}
\cr
&-&
{\cK}_1^{\dot\alpha\alpha}{\widetilde{\cK}}_2^{\dot\beta\beta}
S^+_{\alpha\beta}(Y)S^-_{\dot\alpha\dot\beta}(Z) \, -\,
{\widetilde{\cK}}_1^{\dot\alpha\alpha}
{\cK}_2^{\dot\beta\beta}
S^-_{\alpha\beta}(Y)S^+_{\dot\alpha\dot\beta}(Z)
\biggr] \, C_{\rm norm}
\ , 
\eea
where $i=1\ldots 4$ and $a=1\ldots 4$ label the first 
and second group of four bosonic directions of the 
pp-wave geometry, respectively.
Full details about the expressions appearing in
\eqref{pref} can be found in the original paper
\cite{Panknew} or, for instance, in the review
\cite{Shahin}.
We will only need the following expressions:
\beqa
V_{ij}&=&\delta_{ij}\,
\left[ 1+\frac{1}{12}(Y^4+Z^4)+\frac{1}{144}Y^4Z^4\right]
-\frac{i}{2}\left[Y^2_{ij}(1+\frac{1}{12}Z^4)-
Z^2_{ij}(1+\frac{1}{12}Y^4)\right]
\cr
&+&
\frac{1}{4} (Y^2Z^2)_{ij} \ ,
\\
V_{ab}&=&\delta_{ab}\left[ 1-\frac{1}{12}(Y^4+Z^4)+
\frac{1}{144}Y^4Z^4\right]
-\frac{i}{2}\left[Y^2_{ab}(1-\frac{1}{12}Z^4)-Z^2_{ab}
(1-\frac{1}{12}Y^4)\right]
\cr
&+&\frac{1}{4} (Y^2Z^2)_{ab}
\ ,
\eeqa
where
\beq
Y^{\alpha\beta}\, = \,
\sum_{\rr=1}^3 \sum_{n\geq 0} \,
{\bar G}_{n (\rr)} \, b^{\dagger\alpha\beta}_{n (\rr)}\ , \ \ \ \ \
Z^{\dot\alpha\dot\beta}\, =\,
\sum_{\rr=1}^3 \sum_{n\geq 0}  \, {\bar G}_{n (\rr)}\,
b^{\dagger\dot\alpha\dot\beta}_{-n (\rr)}\ ,
\eeq
\beq
Y^2_{\alpha\beta}\, :=\,  Y_{\alpha\gamma}Y_{\beta}^{\ \gamma}
\ ,  \ \ \ \ \ \ \
Y^4:= Y^2_{\alpha\beta}{(Y^2)}^{\alpha\beta}
\ .
\eeq
Similar expressions hold for the $Z$'s.
The matrices $\bar{G}_{n(\rr)}$ are given in \mbox{(3.12)} of
\cite{Panknew}.
Finally, the overall normalisation $ C_{\rm norm}$
cannot be fixed by imposing the dynamical constraints,
and is determined (once and for all) by requiring agreement
with  a single field theory calculation. 
Its value will be taken
to be:
\bea
\label{ccp}
C_{\rm norm}=- 
\frac{g_2}{2}{1 \over \sqrt{J}}\frac{1}{\sqrt{y(1-y)}}
\ .
\eea

\Appendix{Summing over the BMN phase factors}
We report here the expressions for the coefficients
$P_I $ and $P_{II}$ which arise after summing over
the BMN phase factors in the interacting diagrams
derived in
section \ref{sec-fer}. Defining
\beq
q = e^{2\pi i m /  J} \ , \qquad
q_1 = e^{2\pi i n /  J_1} \ ,
\eeq
the expressions for $P_{I}$ and $P_{II}$ are given by
\beq
\label{fFf}
P_{I} = \sum_{l=0}^{J_1}\  (\bar{q} q_1)^{l}
\ \bar{q} \ , \qquad
P_{II} = \sum_{l=0}^{J_1} \ (\bar{q} q_1)^{l}
\ .
\eeq
We also need to evaluate
the quantity  $2(P_{I} + \bar{P}_{I}) - 2(P_{II} + \bar{P}_{II})$,
which in the BMN limit is
\beqa
\label{totalpf}
2(P_{I} + \bar{P}_{I}) - 2(P_{II} + \bar{P}_{II})
&=&
 - {8 m \over {m - n/  y}}\,  \sin^2 \pi m y
\ .
\eeqa
\Appendix{The functions $X$, $Y$ and $H$}
\label{app-functs}
The expressions for three-point functions of BMN operators with
scalar, vector,  mixed or fermion 
impurities  involve the integral
\beq
\label{X1234}
 X_{1234} = \int d^4z \
\Delta (x_1 - z) \Delta (x_2 - z) \Delta (x_3 - z) \Delta (x_4 - z)
\ \ .
\eeq
$X_{1234}$ develops  a $\log x_{12}^2$ term  $X$
as $x_1$ approaches $x_2$, which repeatedly appears
in section \ref{sec-fer}. The expression for $X$ is
\cite{CKTvec}
\beq
\label{X}
X\ := \ \left. X_{1234}\right|_{x_3 = x_4}\  =  \
{\log \, ( x_{12}\Lambda)^{-1} \over 8 \pi^2 \, (4 \pi^2 x_{31}^2)^2}
\ .
\eeq
Another important function ubiquitously 
appearing in the calculations is
\beq
\label{Ydef}
Y_{123} = \int d^4z \,
\Delta (x_1 - z) \Delta (x_2 - z) \Delta (x_3 - z)
\ .
\eeq
It is easy to realise that, as $x_{12} \to 0$, the function $Y_{123}$
contains a logarithmic term given by
\beq
\label{pp}
\left. Y_{123}\right|_{x_{12} \to 0}  =
 - {1 \over 2^4 \pi^2}
\Delta (x_{13})
 \log x_{12}^2 \
\ .
\eeq
One also needs the following expression for the $ \log x_{12}^2$ term in the
 first derivative of $Y$:
\beq
\label{ii}
(\partial_{1 \a}
Y_{123})_{x_{12} \to 0}  =
 {1 \over 2^5 \pi^2}
 \log x_{12}^2 \
\partial_{3\a} \Delta (x_13)
\ .
\eeq
Notice  that \eqref{ii} should be derived directly
from \eqref{Ydef} rather than
by differentiating  \eqref{pp}.

In the calculation, we also encounter the function
$H$ defined by
\beq
H_{14,23} =
(\partial_{\m}^{x_1} -  \partial_{\m}^{x_4})
(\partial_{\m}^{x_2} -  \partial_{\m}^{x_3})
\int d^4 z \  d^4 t\ \ \D (x_1 - z ) \D (x_4 - z )
  \D (x_2 - t ) \D (x_3 - t )
\D (z - t )
\ ,
\eeq
which can be evaluated with the useful relation  proved in
\cite{BKPSS}
\beq
\label{H}
{H_{14,23} \over \D_{14}\D_{23} }=
X_{1234} \left( {1\over \D_{12}\D_{43} }-{1\over \D_{13}\D_{24} }\right) +
G_{1,23} - G_{4,23}+G_{2,14}-G_{3,14}
\ \ ,
\eeq
where $\D_{ij} = \D (x_i - x_j)$ and
\beq
G_{i,jk}= Y_{ijk}\left( {1 \over \D_{ik}} - {1 \over \D_{ij}}\right)
\ .
\eeq
We can recast \eqref{H} as
\beqa
H_{14,23} &=& -X_{1234} {\Delta_{14}\Delta_{23}\over \Delta_{13}\Delta_{24}}
\ +  \
\left( {Y_{123} \over \D_{13}} +  {Y_{124} \over \D_{24}} \right)
\D_{14}\D_{23} \ + \ \cdots
\nonumber \\ \cr
&= & H_{I} \ + \ H_{II} + \ \cdots
\ \ ,
\eeqa
where the dots stand for terms which either vanish or 
do not contain the
$\log x_{12}^2$.

\Appendix{More detailed calculations for the
evaluation of the Feynman diagrams}
The three-point functions with fermion BMN operators
discussed in section 5 
are expressed in terms of $J_A$, $J_B$ and  $J_C$
defined in \eqref{JA}-\eqref{JC}.
Here we sketch  the calculation of the $\log(x_{12}^2)$ parts
of these quantities. 
Let us start by calculating the following
integral:
\bea
A&=&\partial_k^1\partial_k^4H_{1432}=
\int d^4z \: d^4t \:\partial_k^z\D_{1z}
\:\partial_k^z\D_{4z}\D_{zt}\:\D_{2t}\D_{3t}
=\\ &-&
\int d^4z\: d^4t\:\D_{1z}\Box_z\D_{4z}\:
\D_{zt}\D_{2t}\D_{3t}
\ -\ \int d^4z \: d^4t\:\D_{1z}\partial_k^z\D_{4z}\:\partial_k^z
\D_{zt}\:\D_{2t}\D_{3t}
\ .
\nonumber
\eea
The box acting on the propagator  gives a delta function which 
eliminates the $z$ integration, 
giving a result proportional to $Y_{234}$.
$Y_{234}$, however, does not contain any $\log x_{12}^2$ term
so for our purposes this term can safely be ignored.
Therefore we are left with:
\bea
A&=&
\int  d^4z \: d^4t \:\partial_k^z\D_{1z}\:\D_{4z}\partial_k^z\D_{zt}
\:\D_{2t}\D_{3t}+\int  d^4z \: d^4t \:\D_{1z}\D_{4z}\Box_z\D_{zt}\:
\D_{2t}\D_{3t}\nonumber\\
&=&-\int  d^4z \: d^4t \:\Box_z\D_{1z}\:\D_{4z}\D_{zt}\D_{2t}\D_{3t}-
\int d^4z \: d^4t \:\partial_k^z\D_{1z}
\:\partial_k^z\D_{4z}\D_{zt}\:\D_{2t}\D_{3t}-X_{1234}\nonumber\\
&=&\D_{14}Y_{123}-A-X_{1234}
\ . 
\eea
{}From the last expression one can obtain  $A$:
\bea
A=\frac{1}{2}\bigl(-X_{1234}+\D_{14}Y_{123}\bigr)
\ . 
\eea
In the above derivation ,  
we have integrated by parts with respect to $z$
several times, and we used $\Box_x \D(x)=-\d(x)$.
Since the $\log x_{12}$ terms of $ X_{1234}$ and $Y_{123}$
are well known (see Appendix \ref{app-functs}),
the same is also true for $A$.

Upon using the useful identity $(\partial^1_{\mu}+
\partial^2_{\mu}+\partial^3_{\mu}+\partial^4_{\mu})H_{1423}=0$,
and the expression for $A$ derived above, 
one can evaluate
$ (\partial^3\cdot\partial^4 +\partial^2\cdot\partial^4)H_{1423}$:
\bea
\label{plus}
(\partial^2+\partial^3)\cdot\partial^4H_{1423}
\, =\, -(\partial^1+\partial^4)\cdot\partial^4H_{1423}
\, =\, -A-\Box_4H_{1423} \rightarrow  \, = \, -A
\ , 
\eea
since again $\Box_4$ acting on $H_{1423}$ 
does not give rise to a $\log x_{12}^2 $ term.
One can also evaluate the difference
$(\partial^3\cdot\partial^4 -\partial^2\cdot\partial^4)H_{1423}$
using
\bea
\label{boxes}
\partial^i\cdot\partial^j
H_{1423}\, 
= \, 
\frac{1}{2}(\Box_k+\Box_l-\Box_i-\Box_j)H_{1423}
\,  + \, 
\partial^k\cdot\partial^l H_{1423}
\ , 
\eea
where \eqref{boxes}  holds for $i\neq j\neq k\neq l$.

Starting from
\bea
H_{14,23}\ =\ (\partial^1-\partial^4)\cdot
(\partial^2-\partial^3)H_{1423}
\ , 
\eea
substituting for $\partial^1\cdot\partial^2$ and
 $\partial^1\cdot\partial^3$ the corresponding expressions 
from  \eqref{boxes}, and solving for
$(\partial^3\cdot\partial^4 -\partial^2\cdot\partial^4)H_{1423}$, 
we obtain:
\bea
\label{minus}
(\partial^3\cdot\partial^4 -\partial^2\cdot\partial^4)H_{1423}
\ = \
\frac{1}{2}\Big[ H_{14,23}+(\Box_2-\Box_3)H_{1423}\Big]
\ . 
\eea
Now, since the divergences of the right hand side of \eqref{minus}
are known \cite{GKT},  the divergence of
$(\partial^3\cdot\partial^4 -\partial^2\cdot\partial^4)H_{1423}$
is also known.
In conclusion, we have computed the $\log x_{12}^2 $ parts of
 \eqref{minus} and  \eqref{plus}. That  means we can evaluate
 the $\log x_{12}^2$ parts of 
$\partial^3\cdot\partial^4 \, H_{1423}$ and
$\partial^2\cdot\partial^4 \, H_{1423}$ separately.

We are now in position to write down the expressions for the $J$'s
as functions of $X_{1234}$, $Y_{123}$ and $Y_{124}$.
These are the following:
\bea
\label{6A}
J_A&=&
-\frac{1}{2}(X_{1234}+\D_{41}Y_{123})\, + \cdots \ , 
\\
\label{6B}
J_B&=&
-\frac{1}{2}(-X_{1234}+\D_{23}Y_{124})\,  + \cdots \ , 
\\
\label{6C}
J_C&=&
\frac{1}{4}( -\D_{41}Y_{123}-X_{1234}\, +\D_{23}Y_{124}
+H_{14,23})+ \cdots \ \ , 
\eea
where the dots stand for terms which do not contain the
$\log x_{12}^2$.
$H_{14,23}$ is given in \eqref{H}.
In the evaluation of the diagrams involving 
the compensating term, 
we also made use of the following relations:
\bea
\partial^4_{\nu}\partial^3_{\mu}X_{1234}\mid_{x_3\ = \ 
x_4,x_{12}\rightarrow 0}& =& -\frac{1}{(4\pi^2)^3}
\frac{x_{3\mu}x_{3\nu}}{(x_3^2)^4}\log x_{12}^2
\ , 
\\ \cr
\partial^4_{\nu}\partial^4_{\mu}
X_{1234}\mid_{x_3=x_4,x_{12}\rightarrow 0} & = & 
\D_{23}\partial^4_{\nu}\partial^4_{\mu}
Y_{124}\mid_{x_3=x_4,x_{12}\rightarrow 0}
\\ \nonumber \cr
&=&
\frac{\log x_{12}^2}{2(4\pi^2)^3(x_3^2)^3}\Bigl(
\d_{\mu\nu}
-4\frac{x_{3\mu}x_{3\nu}}{x_3^2}\Bigr)
\ , 
\eea
\bea
\partial^1_{\nu}Y_{123} \mid_{x_3=x_4,x_{12}\rightarrow 0}
&=& 
\frac{-x_{3\nu}}{2^6\pi^4(x_3^2)^2}
\log x_{12}^2
\ , 
\\ \cr
\partial^3_{\nu}Y_{123}
\mid_{x_3=x_4,x_{12}\rightarrow 0}& =& 
\frac{x_{3\nu}}{2^5\pi^4(x_3^2)^2}
\log x_{12}^2
\ . 
\eea



\newpage
\begin{thebibliography}{99}








\bibitem{BMN}
D.~Berenstein, J.~M.~Maldacena and H.~Nastase,
{\it ``Strings in flat space and pp waves from N = 4 super Yang Mills,'' }
JHEP {\bf 0204} (2002) 013,
{\tt hep-th/0202021}.

\bibitem{Metsaev:2001bj}
R.~R.~Metsaev,
 {\it ``Type IIB Green-Schwarz superstring in plane wave Ramond-Ramond
background,''}
Nucl.\ Phys.\ B {\bf 625} (2002) 70,
{\tt hep-th/0112044}.


\bibitem{zanon}
A.~Santambrogio and D.~Zanon,
{\it ``Exact anomalous dimensions of {\cal N}=4 Yang-Mills operators with
large R charge,''} Phys.\ Lett.\ B {\bf 545} (2002) 425,
{\tt hep-th/0206079}.

\bibitem{ver}
H.~Verlinde,
{\it ``Bits, matrices and 1/N,''}
JHEP {\bf 0312} (2003) 052, 
{\tt hep-th/0206059}.


\bibitem{gross2}
D.~J.~Gross, A.~Mikhailov and R.~Roiban,
{\it ``A calculation of the
plane wave string Hamiltonian from N = 4  super-Yang-Mills
theory,''}
JHEP {\bf 0305} (2003) 025, 
{\tt hep-th/0208231}.


\bibitem{KPSS}
C.~Kristjansen, J.~Plefka, G.~W.~Semenoff and M.~Staudacher,
{\it ``A new double-scaling limit of N = 4 super Yang-Mills theory and PP-wave  strings,''}
Nucl.\ Phys.\ B {\bf 643} (2002) 3,
{\tt hep-th/0205033}.

\bibitem{Constable1}
N.~R.~Constable, D.~Z.~Freedman, M.~Headrick, S.~Minwalla, L.~Motl,
A.~Postnikov and W.~Skiba,
{\it ``PP-wave string interactions from perturbative Yang-Mills theory,''  }
JHEP {\bf 0207} (2002) 017, {\tt hep-th/0205089}.

\bibitem{Gomis}
J.~Gomis, S.~Moriyama and J.~w.~Park,
{\it ``SYM description of SFT Hamiltonian in a pp-wave
background,''}
Nucl.\ Phys.\ B {\bf 659} (2003) 179,
{\tt hep-th/0210153.}

\bibitem{Gomis2}
J.~Gomis, S.~Moriyama and J.~w.~Park,
{\it ``SYM description of pp-wave string interactions: Singlet sector and
arbitrary impurities,''}
Nucl.\ Phys.\ B {\bf 665} (2003) 49,
{\tt hep-th/0301250.}


\bibitem{GKT}
G.~Georgiou, V.~V.~Khoze and G.~Travaglini,
{\it ``New tests of the pp-wave correspondence,''}
JHEP {\bf 0310} (2003) 049,
{\tt hep-th/0306234}.





\bibitem{SV1}
M.~Spradlin and A.~Volovich,
{\it ``Superstring interactions in a pp-wave background,''}
Phys.\ Rev.\ D {\bf 66} (2002) 086004,
{\tt hep-th/0204146}.




\bibitem{SV2}
M.~Spradlin and A.~Volovich,
{\it ``Superstring interactions in a pp-wave background. II,''}
JHEP {\bf 0301} (2003) 036,
{\tt hep-th/0206073}.

\bibitem{AP}
A.~Pankiewicz,
{\it ``More comments on superstring interactions in the pp-wave background,''}
JHEP {\bf 0209} (2002) 056,
{\tt hep-th/0208209}.



\bibitem{PS}
A.~Pankiewicz and B.~Stefanski,
{\it ``pp-wave light-cone superstring field theory,''}
Nucl.\ Phys.\ B {\bf 657} (2003) 79,
{\tt hep-th/0210246}.


\bibitem{KKLP}
Y.~j.~Kiem, Y.~b.~Kim, S.~m.~Lee and J.~m.~Park,
{\it ``pp-wave / Yang-Mills correspondence: An explicit check,''}
Nucl.\ Phys.\ B {\bf 642} (2002) 389,
{\tt hep-th/0205279}.

\bibitem{LMP}
P.~Lee, S.~Moriyama and J.~w.~Park,
{\it ``Cubic interactions in pp-wave light cone string field theory,''}
Phys.\ Rev.\ D {\bf 66} (2002) 085021,
{\tt hep-th/0206065}.


\bibitem{CKPRT}
C.~S.~Chu, V.~V.~Khoze, M.~Petrini, R.~Russo and A.~Tanzini,
{\it ``A note on string interaction on the pp-wave background,''}
{\tt hep-th/0208148}.


\bibitem{KKPR}
Y.~j.~Kiem, Y.~b.~Kim, J.~Park and C.~Ryou,
{\it ``Chiral primary cubic interactions from pp-wave supergravity,''}
JHEP {\bf 0301} (2003) 026,
{\tt hep-th/0211217}.


\bibitem{bits3}
J.~Pearson, M.~Spradlin, D.~Vaman, H.~Verlinde and A.~Volovich,
{\it ``Tracing the string: BMN correspondence at finite $J^2/N$,''}
JHEP {\bf 0305} (2003) 022,
{\tt hep-th/0210102}.

\bibitem{CKTvec}
C.~S.~Chu, V.~V.~Khoze and G.~Travaglini,
{\it ``BMN operators with vector impurities,
$\bb{Z}_2$ symmetry and pp-waves,''}
JHEP {\bf 0306} (2003) 050,
{\tt hep-th/0303107}.



\bibitem{Eden:2003sj}
B.~Eden,
{\it ``On two fermion BMN operators,''}
Nucl.\ Phys.\ B {\bf 681} (2004) 195,
{\tt hep-th/0307081}.

\bibitem{Bianchi:2003eg}
M.~Bianchi, G.~Rossi and Y.~S.~Stanev,
{\it ``Surprises from the resolution of operator mixing in N = 4 SYM,''}
{\tt hep-th/0312228}.





\bibitem{Panknew}
A.~Pankiewicz,
{\it ``An alternative formulation of light-cone string field theory on the  plane wave,''}
JHEP {\bf 0306} (2003) 047,
{\tt hep-th/0304232}.


\bibitem{Pankiewicz:2003ap}
A.~Pankiewicz and B.~J.~Stefanski,
{\it ``On the uniqueness of plane-wave string field theory,''}
{\tt hep-th/0308062.}



\bibitem{Gomis3}
J.~Gomis, S.~Moriyama and J.~w.~Park,
{\it ``Open + closed string field theory from gauge fields,''},
Nucl.\ Phys.\ B {\bf 678} (2004) 101,
{\tt hep-th/0305264}.




\bibitem{DiVecchia:2003yp}
P.~Di Vecchia, J.~L.~Petersen, M.~Petrini, R.~Russo and A.~Tanzini,
{\it ``The 3-string vertex and the AdS/CFT duality in the pp-wave limit,''}
{\tt hep-th/0304025.}


\bibitem{Slansky:yr}
R.~Slansky,
{\it ``Group Theory For Unified Model Building,''}
Phys.\ Rept.\  {\bf 79} (1981) 1.


\bibitem{Parnachev:2002kk}
A.~Parnachev and A.~V.~Ryzhov,
{``Strings in the near plane wave background and AdS/CFT,''}
JHEP {\bf 0210} (2002) 066,
{\tt hep-th/0208010}.


\bibitem{gursoy}
U.~Gursoy,
{\it ``Vector operators in the BMN correspondence,''}
JHEP {\bf 0307} (2003) 048,
{\tt hep-th/0208041}.


\bibitem{beisert}
N.~Beisert,
{\it ``BMN operators and superconformal symmetry,''}
Nucl.\ Phys.\ B {\bf 659} (2003) 79,
{\tt hep-th/0211032}.




\bibitem{BKPSS}
N.~Beisert, C.~Kristjansen, J.~Plefka, 
G.~W.~Semenoff and M.~Staudacher,
{\it ``BMN correlators and operator mixing in N = 4 super Yang-Mills theory,''}
Nucl.\ Phys.\ B {\bf 650} (2003) 125,
{\tt hep-th/0208178}.

\bibitem{CKT}
C.~S.~Chu, V.~V.~Khoze and G.~Travaglini,
{\it ``Three-point functions in N = 4 Yang-Mills theory and pp-waves,''  }
JHEP {\bf 0206} (2002) 011,  {\tt hep-th/0206005}.


\bibitem{CKT2}
C.~S.~Chu, V.~V.~Khoze and G.~Travaglini,
{\it ``pp-wave string interactions from n-point correlators of BMN operators,''}
JHEP {\bf 0209} (2002) 054, 
{\tt hep-th/0206167}.

\bibitem{Constable2}
N.~R.~Constable, D.~Z.~Freedman, M.~Headrick and S.~Minwalla,
{\it ``Operator mixing and the BMN correspondence,''}
JHEP {\bf 0210} (2002) 068,
{\tt hep-th/0209002}.



\bibitem{Shahin}
D.~Sadri and M.~M.~Sheikh-Jabbari,
{\it ``The plane-wave / super Yang-Mills duality,''}
{\tt hep-th/0310119.}


\bibitem{Fradkin:is}
E.~S.~Fradkin and M.~Y.~Palchik,
{\it ``Conformal Quantum Field Theory In D-Dimensions,''}
Mathematics and its applications, {\bf 376},
Kluver, Dordrecht  1996.


\bibitem{Janik}
R.~A.~Janik,
{\it ``BMN operators and string field theory,''}
Phys.\ Lett.\ B {\bf 549} (2002) 237,
{\tt hep-th/0209263}.


\bibitem{Minahan:2002ve}
J.~A.~Minahan and K.~Zarembo,
{\it ``The Bethe-ansatz for N = 4 super Yang-Mills,''}
JHEP {\bf 0303} (2003) 013, 
{\tt hep-th/0212208}.


\bibitem{B1}
N.~Beisert, C.~Kristjansen, J.~Plefka and M.~Staudacher,
{\it ``BMN gauge theory as a quantum mechanical system,''}
Phys.\ Lett.\ B {\bf 558} (2003) 229,
{\tt hep-th/0212269}.

\bibitem{SV3}
M.~Spradlin and A.~Volovich,
{\it ``Note on plane wave quantum mechanics,''}
Phys.\ Lett.\ B {\bf 565} (2003) 253,
{\tt hep-th/0303220}.

\bibitem{B2}
N.~Beisert, C.~Kristjansen and M.~Staudacher,
{\it ``The dilatation operator of N = 4 super Yang-Mills theory,''}
{\tt hep-th/0303060}.

\bibitem{Beisert:2003jj}
N.~Beisert,
{\it ``The complete one-loop dilatation operator of N = 4 super Yang-Mills
theory,''}
Nucl.\ Phys.\ B {\bf 676} (2004) 3, 
{\tt hep-th/0307015}.



\bibitem{Beisert:2003yb}
N.~Beisert and M.~Staudacher,
{\it ``The N = 4 SYM integrable super spin chain,''}
Nucl.\ Phys.\ B {\bf 670} (2003) 439, 
{\tt hep-th/0307042}.

\bibitem{Beisert:2003ys}
N.~Beisert,
{\it ``The su(2$|$3) dynamic spin chain,''}
{\tt hep-th/0310252}.



\bibitem{Kristjansen:2004ei}
C.~Kristjansen,
{\it ``Three-spin strings on AdS(5) x S**5 from N = 4 SYM,''}
{\tt hep-th/0402033.}


\bibitem{Arutyunov:2004xy}
G.~Arutyunov and M.~Staudacher,
{\it ``Two-loop commuting charges and the string / gauge duality,''}
{\tt hep-th/0403077}.



\bibitem{Mandelstam:hk}
S.~Mandelstam,
{\it ``Interacting String Picture Of The Neveu-Schwarz-Ramond Model,''}
Nucl.\ Phys.\ B {\bf 69} (1974) 77.


\bibitem{Green:1982tc}
M.~B.~Green and J.~H.~Schwarz,
{\it ``Superstring Interactions,''}
Nucl.\ Phys.\ B {\bf 218} (1983) 43.

\bibitem{Green:hw}
M.~B.~Green, J.~H.~Schwarz and L.~Brink,
{\it ``Superfield Theory Of Type Ii Superstrings,''}
Nucl.\ Phys.\ B {\bf 219} (1983) 437.

\bibitem{Green:fu}
M.~B.~Green and J.~H.~Schwarz,
{\it ``Superstring Field Theory,''}
Nucl.\ Phys.\ B {\bf 243} (1984) 475.




\bibitem{gursoy2}
U.~Gursoy,
{\it ``Predictions for pp-wave string amplitudes from perturbative SYM,''}
JHEP {\bf 0310} (2003) 027,
{\tt hep-th/0212118}.



\bibitem{klose}
T.~Klose,
{\it ``Conformal dimensions of two-derivative BMN operators,''}
JHEP {\bf 0303} (2003) 012,
{\tt hep-th/0301150}.



\bibitem{GK}
G.~Georgiou and V.~V.~Khoze,
{\it ``BMN operators with three scalar impurities and the
  vertex-correlator  duality in pp-wave,''}
JHEP {\bf 0304} (2003) 015,
{\tt hep-th/0302064}.


\bibitem{Bianchi}
M.~Bianchi, B.~Eden, G.~Rossi and Y.~S.~Stanev,
{\it ``On operator mixing in N = 4 SYM,''}
Nucl.\ Phys.\ B {\bf 646} (2002) 69,
{\tt hep-th/0205321}.


\bibitem{MT}
R.~R.~Metsaev and A.~A.~Tseytlin,
 {\it ``Exactly solvable model of superstring in plane wave Ramond-Ramond
background,''}
Phys.\ Rev.\ D {\bf 65} (2002) 126004, 
{\tt hep-th/0202109}.


\bibitem{'tHooft:fv}
G.~'t Hooft,
 {\it ``Computation Of The Quantum Effects Due To A Four-Dimensional
Pseudoparticle,''}
Phys.\ Rev.\ D {\bf 14} (1976) 3432
[Erratum-ibid.\ D {\bf 18} (1978) 2199].

\bibitem{HSSV}
Y.~H.~He, J.~H.~Schwarz, M.~Spradlin and A.~Volovich,
{\it ``Explicit formulas for Neumann coefficients in the plane-wave geometry,''}
Phys.\ Rev.\ D {\bf 67} (2003) 086005,
{\tt hep-th/0211198}.

\bibitem{RR}
C.~S.~Chu, M.~Petrini, R.~Russo and A.~Tanzini,
{\it ``String interactions and discrete symmetries of the pp-wave background,''}
Class.\ Quant.\ Grav.\  {\bf 20} (2003) S457, 
{\tt hep-th/0211188}.


\bibitem{Lucietti:2004wy}
J.~Lucietti, S.~Schafer-Nameki and A.~Sinha,
{\it ``On the plane-wave cubic vertex,''}
{\tt hep-th/0402185}.


\bibitem{CK}
C.~S.~Chu and V.~V.~Khoze,
{\it ``Correspondence between the 3-point BMN correlators
 and the  3-string  vertex on the pp-wave,''}
JHEP {\bf 0304} (2003) 014,
{\tt hep-th/0301036}.





































\end{thebibliography}
\end{document}